\documentclass[preprintnumbers,amssymb,amsmath,nofootinbib,12pt]{revtex4}
\usepackage[dvips]{graphicx}
\usepackage{dcolumn}
\usepackage{bm}
\usepackage{epsfig,amsmath,amssymb,verbatim,mathrsfs,array,layout,textcomp,amssymb,latexsym,slashed}
\input epsf.tex
\newcommand{\beq}{\begin{eqnarray}}
\newcommand{\eeq}{\end{eqnarray}}

\def\ltap{\ \raise.3ex\hbox{$<$\kern-.75em\lower1ex\hbox{$\sim$}}\ }
\def\gtap{\ \raise.3ex\hbox{$>$\kern-.75em\lower1ex\hbox{$\sim$}}\ }
\def\CO{{\cal O}}
\def\CL{{\cal L}}

\def\CO{{\cal O}}
\def\CL{{\cal L}}

\def\eg{{\it e.g.}}

\def\be{\begin{equation}}
\def\ee{\end{equation}}
\def\bea{\begin{eqnarray}}
\def\eea{\end{eqnarray}}

\newcommand{\hc}{{\rm h.c.}}
\newcommand{\half}{\frac{1}{2}}

\newcommand{\vckm}{V^{\rm CKM}}
\newcommand{\rqmat}{\mathrm{diag}(1,1,r_Q)}
\newcommand{\rqsmat}{\mathrm{diag}(1,1,r_Q^2)}
\newcommand{\rdmat}{\mathrm{diag}(1,1,r_D)}
\newcommand{\rdsmat}{\mathrm{diag}(1,1,r_D^2)}

\def\mysection#1{{{\bf #1}.~}}

\begin{document}

\title{Ultra Visible Warped Model From Flavor Triviality \\
\& Improved Naturalness}

\author{C\'edric Delaunay}
\affiliation{Department of Particle Physics and Astrophysics,
Weizmann Institute of Science, Rehovot 76100, Israel}
\author{Oram Gedalia}
\affiliation{Department of Particle Physics and Astrophysics,
Weizmann Institute of Science, Rehovot 76100, Israel}
\author{Seung J. Lee}
\affiliation{Department of Particle Physics and Astrophysics,
Weizmann Institute of Science, Rehovot 76100, Israel}
\author{Gilad Perez}
\affiliation{Department of Particle Physics and Astrophysics,
Weizmann Institute of Science, Rehovot 76100, Israel}
\author{Eduardo Pont\'on }
\affiliation{Department of Physics, Columbia University, 538 W. 120th St, New York, NY 10027, USA}

\vskip .05in

\begin{abstract}
\vskip .05in

A warped extra-dimensional model, where the Standard Model
Yukawa hierarchy is set by UV physics, is shown to have a sweet
spot of parameters with improved experimental visibility and
possibly naturalness. Upon marginalizing over all the model
parameters, a Kaluza-Klein scale of 2.1~TeV can be obtained at
2$\sigma$ ($95.4\%$ CL) without conflicting with electroweak
precision measurements. Fitting all relevant parameters
simultaneously can relax this bound to 1.7~TeV. In this bulk
version of the Rattazzi-Zaffaroni shining model, flavor
violation is also highly suppressed, yielding a bound of
2.4~TeV. Non-trivial flavor physics at the LHC in the form of
flavor gauge bosons is predicted. The model is also
characterized by a depletion of the third generation
couplings~-- as predicted by the general minimal flavor
violation framework~-- which can be tested via flavor precision
measurements. In particular, sizable CP violation in $\Delta
B=2$ transitions can be obtained, and there is a natural region
where $B_s$ mixing is predicted to be larger than $B_d\,$
mixing, as favored by recent Tevatron data. Unlike other
proposals, the new contributions are not linked to Higgs or any
scalar exchange processes.

\end{abstract}

\maketitle
\section{Introduction}

Plunging the Standard Model (SM) in a warped extra-dimension
provides new perspectives on understanding electroweak symmetry
breaking (EWSB), offering a new way to solve the gauge
hierarchy problem~\cite{Randall:1999ee}. The Randall-Sundrum
(RS) class of models also offers a simple way to address the SM
flavor puzzle by localizing the SM fermions away from the Higgs
vacuum expectation value (VEV) with $\mathcal{O}(1)$
parameters~\cite{ArkaniHamed:1999dc, Grossman:1999ra}, which is
referred to as the anarchic approach. In addition, the anarchic
setup protects against large flavor and CP violation via the so
called RS-GIM mechanism~\cite{Agashe:2004ay, Agashe:2004cp,
Cacciapaglia:2007fw}. Yet, a residual little CP problem, in the
form of too large contributions to
$\epsilon_K$~\cite{Bona:2007vi, Davidson:2007si, Csaki:2008zd,
Blanke:2008zb,Bauer:2009cf} and electric dipole
moments~\cite{Agashe:2004ay, Agashe:2004cp, Cheung:2007bu},
remains. (Some more RS flavor issues can be found in {\it
e.g.}~\cite{Huber:2003tu, Burdman:2003nt,Moreau:2006np,
Agashe:2006wa, Casagrande:2008hr,
Agashe:2008uz,Blanke:2008yr,Csaki:2009bb,
Buras:2009ka,Gedalia:2009ws}.) Furthermore, this framework
calls for improvement on naturalness since a fine-tuning of
worse than ${\cal O }(10\%)$~\cite{Agashe:2004rs,
Agashe:2005dk, Csaki:2008zd,Panico:2008bx} of the electroweak
(EW) scale is required to comply with EW precision tests
(EWPTs)~\cite{Agashe:2003zs,Carena:2006bn}. In the best of all
known RS models, including a custodial symmetry for $Z\to b\bar
b$, the lore is that this pushes the Kaluza-Klein (KK) scale
above 3 TeV (below we argue that these numbers may be too
optimistic).

It has been known for some time that changing the position of
the light fermions, thus giving up on the virtues of the
anarchic approach, may result in a better EW fit. In
particular, if the profile of all the light fermions is close
to being flat, a suppression of the Peskin-Takeuchi $S$
parameter is obtained~\cite{Agashe:2003zs, Cacciapaglia:2004jz,
Foadi:2004ps, Chivukula:2005bn, Dawson:2007yk,
Accomando:2008jh, Dawson:2008as}. This would allow to lower the
KK scale and possibly improve the naturalness of the model. It
is interesting that such a fermion setup is consistent with
imposing in the bulk the approximate SM flavor symmetries:
$U(2)_Q\times U(2)_U\times U(2)_D\times U(3)_L\times U(3)_E$,
where $Q,U,D$ ($L,E$) correspond to the SM quark (lepton)
doublet, up and down type quark (charged lepton) singlets,
respectively.

In the following, we propose to give up on the warped extra
dimensional built-in mechanism for solving the flavor puzzle
and the RS-GIM protection; after all, no experimental evidence
implies that the flavor hierarchies arise from TeV scale
physics, while, on the other hand, the hierarchy problem does
inevitably point to it. We assume that the Yukawa hierarchy is
set by some unknown physics on the UV brane, while both the
bulk and the IR brane are invariant under the (now gauged) SM
flavor symmetries.\footnote{The lepton symmetry can be also be
broken down to products of $U(2)$. However, for simplicity we
do not consider this possibility nor do we focus on lepton
flavor violation, which is suppressed in our framework, or
neutrino masses. Both issues are discussed
in~\cite{Perez:2008ee, Csaki:2008qq}.} Then the hierarchical
five dimensional (5D) fundamental Yukawa couplings are shined
through the bulk by scalar flavon fields, thus realizing the
approximate SM flavor symmetry structure.

Such a setup was first proposed by Rattazzi and Zaffaroni
(RZ)~\cite{Rattazzi:2000hs}, where the SM fields were localized
on the IR brane as in the original RS1
model~\cite{Randall:1999ee}. In this case, higher-dimensional
operators, which generically contribute to EWPTs and flavor
changing neutral currents (FCNCs), can be suppressed, but only
at the expense of a severe little hierarchy problem.

We show below that a bulk version of the RZ model, with a bulk
Higgs~\cite{Davoudiasl:2000wi}, leads to a very exciting class
of models, where improved agreement with EWPTs is obtained. We
perform a global fit to the EW precision observables,
evaluating the contributions to $S,T$ and $Z\to b \bar b$ at
one-loop order (which are calculable in this model). In order
to compare our model to the celebrated anarchic case, we also
repeat the fit for this case. (However, for simplicity, we only
consider one possible custodial assignment.) As a result of our
analysis, we find a sweet spot in the parameter space of the
bulk RZ model, which allows for a significantly lower KK scale,
such that it would be much easier to observe (or
exclude~\cite{Davoudiasl:2007wf}) at the LHC. Furthermore, we
show that the inclusion of the one-loop contributions to the EW
observables raises the KK scale of the anarchic case. In
addition, the fine-tuning associated with our model is
ameliorated relative to the anarchic case.

The above scenario offers also some interesting perspective on
flavor physics. First of all, the $\epsilon_K$ RS problem is
solved, so that the bound from flavor is considerably weakened.
Second, the model is characterized by a depletion of the third
generation couplings, as predicted by the general minimal
flavor violation framework~\cite{Kagan:2009bn}. The model also
yields sizable CP violation (CPV) in $\Delta B=2$ transitions
with, in particular, the possibility to obtain CPV
contributions in $B_s$ mixing larger than in $B_d\,$, as seems
favored by the Tevatron data at present~\cite{Ligeti:2010ia}.
This is achieved without invoking Higgs or other scalar
exchange processes~\cite{Dobrescu:2010rh,Buras:2010mh,
Jung:2010ik}. Finally, since the bulk flavor symmetry is
gauged, such that large breaking effects from quantum gravity
are avoided, flavor gauge bosons are awaited around the TeV
scale. Such states may be discovered at the LHC~\cite{FGB}.

In short, the main differences between our study and previous
ones are:
\begin{itemize}
\item We give a rational and an explicit model (a bulk RZ
    setup with  some rough speculations on a possible
    extension to grand unification) where the light fermion
    profiles are roughly flat.
\item We choose a custodial representation for the leptons,
    which turns out to significantly improve the result of
    the global EW fit.
\item We emphasize, by calculating explicitly (and via 5D
    power counting), that in the bulk Higgs case the $S$
    parameter is one-loop finite. Furthermore, our
    estimation of the UV sensitive contribution to $S$,
    based on naive dimensional analysis (NDA), shows that
    they are subdominant (for related discussions,
    see~\cite{Carena:2006bn, Carena:2007ua,Dawson:2008as}).
    Thus, the resulting value of $S$ is dominated by finite
    contributions, and is under control.
\item  We use updated input parameters for our EW fit taken
    from~\cite{PDG2010}. Additionally, an appropriate top
    5D Yukawa value, matched to the top mass at the
    relevant scale, is used, and our 5D gauge couplings are
    matched at one-loop.
\item Our statistical treatment consists of two different
    analyses. In the first one we report a bound on the KK
    scale upon marginalizing over all the other model
    parameters, while in the second we produce a bound when
    all the relevant parameters are combined in a
    multi-dimensional fit.

\item Finally, even though for simplicity we have not
    considered the case where the Higgs is a
    pseudo-Goldstone boson (PGB), we provide a rough
    speculation on the fine-tuning of PGB extensions that
    include the above improvements, which can be compared
    to other genuine PGB studies~\cite{Agashe:2004rs,
    Csaki:2008zd, Panico:2008bx}.
\end{itemize}

The rest of the paper is organized as follows. In
Sec.~\ref{sec:model} we describe the warped 5D setup and define
our notation. Then, in Sec.~\ref{sec:ewpt}, the constraints
from EWPTs on this class of models are presented, while  in
Sec.~\ref{sec:flavor} we elaborate on their flavor
phenomenology. Finally, Sec.~\ref{sec:conc} gathers our
conclusions and discusses prospects at the LHC.

\section{The model} \label{sec:model}

We work in a slice of AdS$_5$ space-time. The metric is
$ds^2=(kz)^{-2}\left(\eta_{\mu\nu}dx^\mu dx^\nu-dz^2\right)$
with $\eta_{\mu\nu}={\rm diag}(+---)$ and a curvature scale
$k\simeq 2.4\times 10^{18}\,$GeV, hence solving the hierarchy
problem all the way up to the Planck scale. The slice is
bounded by two branes at $z=R\sim k^{-1}$ and
$z=R\,'\sim\,$TeV$^{-1}$ usually referred to as the UV and IR
branes, respectively. We impose a SU(2)$_L\times
$SU(2)$_R\times $U(1)$_X$ gauge symmetry in the bulk. For
simplicity, in this study we assume that the Higgs field~--
$H\sim (2,2)_0$ under the $(L,R)_X$ custodial gauge group~-- is
a bulk field with VEV $\langle
H\rangle=v_5(z,\beta)/\sqrt{2}\,$, where $v_5(z,\beta)\simeq
vR'/R^{3/2}\sqrt{2(1+\beta)}(z/R')^{2+\beta}$ and $v\simeq
246\,$GeV~\cite{Cacciapaglia:2006mz}. The $\beta$ parameter
sets the VEV localization in the bulk, with $\beta=0$
corresponding to gauge-Higgs unified
models~\cite{Agashe:2004rs, Agashe:2006at}. Eventually, the
model should be lifted to one where the Higgs is realized as a
pseudo-Goldstone boson~\cite{Agashe:2004rs, Contino:2003ve}, so
that the quadratically divergent corrections to the Higgs mass
are cut at the KK scale.\footnote{We leave this specific
analysis to future work.} Therefore, we choose $\beta=0$ in the
following, so we expect our conclusions to approximately hold
also in models where the Higgs is a pseudo-Goldstone boson. We
also gauge in the bulk the non-Abelian part of the SM flavor
symmetry SU(3)$_Q\times $SU(3)$_U\times $SU(3)$_D\times
$SU(3)$_L\times $SU(3)$_E$, such that all flavor changing
effects are controlled by the SM Yukawas, thus realizing the
minimal flavor violation (MFV)
ansatz~\cite{Rattazzi:2000hs,Fitzpatrick:2007sa,Csaki:2009wc}.
The fermions are embedded as $Q\sim(2,2)_{2/3}$,
$U\sim(1,1)_{2/3}$, $D\sim(1,3)_{2/3}\oplus (3,1)_{2/3}$ and
$L\sim(2,2)_{-1}$, $E\sim(1,3)_{0}\oplus(3,1)_{0}$, so they
transform covariantly under the custodial
parity~\cite{Agashe:2006at}.

The bulk gauge symmetry breaks down to the SM gauge group on
the UV brane and still preserves a custodial $SU(2)_{L+R}$
after EWSB, so the $T$ parameter is protected from large bulk
cutoff corrections. The breaking of the flavor group occurs
only on the UV brane, and is shined toward the IR by some
flavon scalar fields $\Phi$, with VEV $\langle \Phi
\rangle\propto Y_I\,$, where $Y_I$ are the 5D Yukawa matrices
($I = U,D,E$). In contrast with most previous studies, we take
the 5D Yukawas to display the hierarchy observed in 4D, which
boils down to assuming that the latter are set by unspecified
UV physics. The large top Yukawa implies a shift in the third
generation bulk masses, while the 5D bottom Yukawa is free to
be taken either large or small. The latter can be regarded as
the large or small $\tan \beta$ cases in two Higgs doublet
models, such as supersymmetric theories. For simplicity we
shall assume in the EW global fit that the 5D bottom Yukawa is
small, and leave the implications of a large bottom Yukawa
option to the flavor physics discussion in
Sec.~\ref{sec:flavor}. This setup guarantees that at low
energies the model belongs to the MFV
framework~\cite{Chivukula:1987py,Hall:1990ac,Gabrielli:1994ff,
Ali:1999we,Buras:2000dm,D'Ambrosio:2002ex,Buras:2003jf,
Buras:2005xt,Hurth:2008jc,Isidori:2009px}, where harmless top
Yukawa resummation is expected and may be observable in the
future. Note that although taking a somewhat larger 5D bottom
Yukawa (but still suppressed compared to the 5D top Yukawa)
would not strongly affect the EWPTs, it would lead to a richer
flavor phenomenology. In addition, flavor violation from the
presence of flavor gauge bosons is also expected, but yet
again, it is going to be subject to MFV protection~\cite{FGB}.
In the following we discuss in more detail the EW and flavor
sectors of our model and their phenomenological implications.

\section{Electroweak precision tests} \label{sec:ewpt}

Models of new physics for the EW scale are tightly constrained,
at the {\it per mile} level, by the measurements at SLD and
LEP, both at and above the $Z$ pole~\cite{:2003ih,:2005ema}, as
well as by the Tevatron. In a large set of such models, the
gauge sector observables, described by the so-called oblique
parameters, capture most (if not all) of the constraints on new
physics. Moreover, the large coupling of the top to the EWSB
sector typically implies sizable non-oblique corrections for
the third generation quarks, notably to the $Z\bar{b}_L b_L$
coupling. The oblique parameters, along with the $Z$ partial
decay width into $b\bar{b}$, constitute a reduced set of EW
precision observables (EWPOs) often sufficient to constrain RS
models~\cite{Agashe:2003zs,Agashe:2005dk, Carena:2006bn,
Bouchart:2008vp,Davoudiasl:2009cd, Bouchart:2009vq,
Casagrande:2010si}. Indeed, whenever localized toward the UV
brane, the (elementary) light fermions are barely sensitive to
EWSB in the IR, and induce negligible corrections to the EWPOs.
In contrast, since the light fermions are  more composite in
our setup, additional non-oblique corrections are expected to
be generated. This requires a more careful study of other
observables, such as the hadronic $Z$ decay width and
observables sensitive to four-fermion operators in the lepton
sector, like atomic parity violation (APV) in heavy nuclei. In
such a highly non-universal new physics model, this implies
that one must look at more than $\mathcal{O}(35)$ EWPOs in
order to assess whether EWPTs are passed. In the following, we
discuss in detail how such a fit is performed. Then, we report
the resulting bounds on the KK scale and estimate the
fine-tuning in our model (as well as in the anarchic case) by
computing the sensitivity of the new physics scale to the input
parameters.

\subsection{Global Fit to Electroweak Observables}

In order to properly include all possible correlations among
the various observables, we perform a global fit to the EW
precision data following the approach of~\cite{Han:2004az,
Han:2005pr}. To do so, we match the relevant dimension-six
operators in the SM to our RS setup
(see~\cite{Davoudiasl:2009cd} for a review), including the most
important, top (and eventually bottom) Yukawa enhanced,
radiative corrections to the $S$ and $T$ parameters and the
$Z\bar{b}_L b_L$ vertex. Radiative corrections to lighter
fermion-gauge boson couplings and to four-fermion operators
will be suppressed by smaller Yukawas. Note that when the
bottom 5D Yukawa is $\mathcal{O}(1)$ or larger, additional loop
contributions to the $Z\bar{b}_L b_L$ vertex involving neutral
currents become important. We have not included such
contributions and therefore will limit our analysis to a
relatively small bottom Yukawa, where these radiative
corrections are subdominant.\footnote{In practice,  this
implies a hierarchy of $\sim 4-5$ between the 5D bottom Yukawa
and the best fit value obtained for the top one.}

\begin{figure}[h!bt]
	\centering
\begin{tabular}{m{5cm} m{5cm} m{5cm}}\includegraphics[scale=0.5]{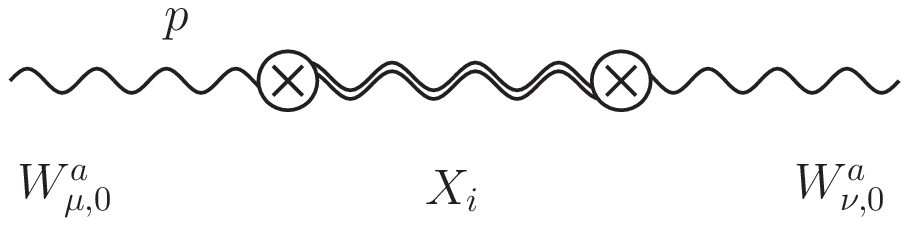} &
\includegraphics[scale=0.5]{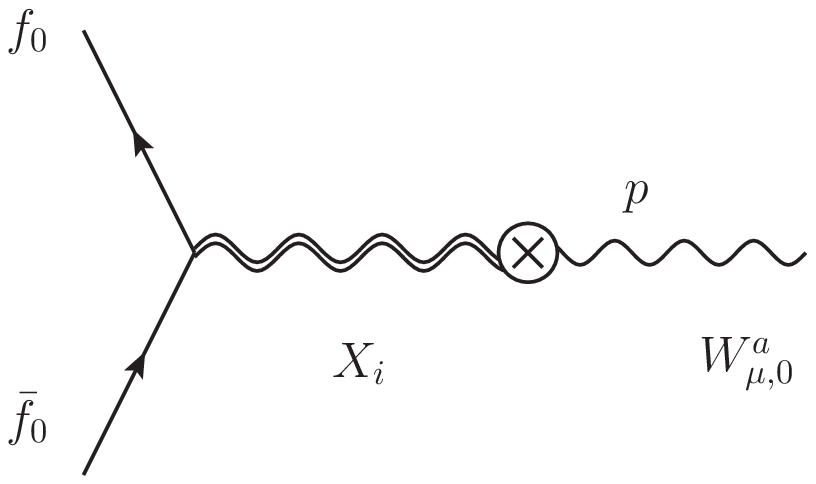} & \includegraphics[scale=0.5]{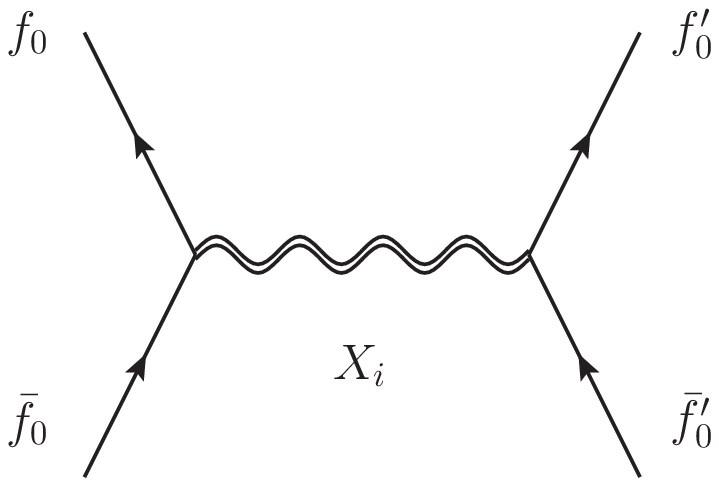}\end{tabular}
	\caption{Tree diagrams contributing at leading order to the EWPO.
	The double line denotes a sum over the various gauge KK-states, while
    the cross represents KK/zero-mode mixing from the Higgs VEV. $W^a_{0}$
    are the SM zero-modes with $a=0,\dots,3$ and $W^0\equiv B$ is the hypercharge gauge field.}
	\label{fig:diagtree}
\end{figure}

We include the following RS contributions to the SM
dimension-six operators. First of all, working at leading order
in $\left(v/m_{\rm KK}\right)^2\ll 1$, the tree-level effects
arise from exchange of KK-gauge bosons through the diagrams of
Fig.~\ref{fig:diagtree}. Additional tree-level contributions
from the left-handed (LH) bottom sector (potentially,
controlled by the top Yukawa coupling) are absent due to the
custodial protection~\cite{Agashe:2006at} (more generally, the
LH down-type sector of the three generations enjoys a custodial
protection). For the up-type, as well as the right-handed down
quark sectors, which are not protected by this symmetry, the
effects are suppressed by the assumed hierarchical nature of
the 5D Yukawa couplings (except for the top which is not, at
present, experimentally constrained).

\begin{figure}[h!bt]
	\centering
\includegraphics[scale=0.5]{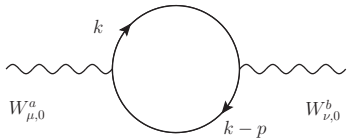}
	\caption{Diagram contributing to the SM gauge boson propagators
    at one-loop.    }
	\label{fig:diagoneloop}
\end{figure}

Furthermore, it is known that isospin breaking in the fermionic
sector leads to sizable corrections to $T$ at
one-loop~\cite{Agashe:2003zs}. This correction is often
negative as a result of the choice of custodial
representations, unless the singlet contribution dominates, in
which case $T$ can be positive at one-loop~\cite{Carena:2006bn,
Carena:2007ua}. On the other hand, the one-loop corrections to
$S$ tend to be positive and relatively small in RS for a
reasonable range of parameters~\cite{Carena:2006bn,
Carena:2007ua}. To prevent the appearance of a large $S$
parameter at tree-level and cancel the effect on the global fit
of the small positive one-loop correction, we will focus on a
region where the light fermions are almost
flat~\cite{Agashe:2003zs}. Notice that since $S$ is not
protected by any symmetry, it could {\it a priori} be UV
sensitive in 5D, whereas $T$ is finite to all orders in
perturbation thanks to the custodial symmetry. However, we show
below that for a bulk Higgs the $S$ parameter is one-loop
finite. Thus the one-loop shifts in {\it both} $S$ and $T$ are
calculable and dominated by the first KK-states. In practice,
we include the first two KK-levels in the fit; higher KK-levels
would yield at most a $(m_{\rm KK}^{(1)}/m_{\rm
KK}^{(3)})^2\sim 10\%$ correction, which we choose to neglect.
Moreover, third generation KK-quarks dominate the shift to the
weak gauge boson two-point functions through the diagram of
Fig.~\ref{fig:diagoneloop}, while other contributions are
suppressed by either gauge couplings or smaller 5D Yukawas.

\begin{figure}[h!bt]
	\centering
\begin{tabular}{ccc}\includegraphics[scale=0.5]{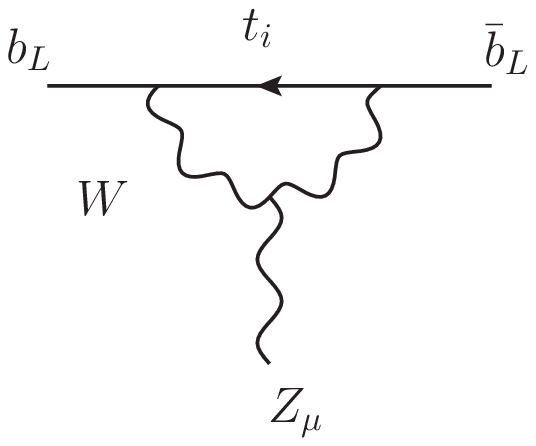} &
\includegraphics[scale=0.5]{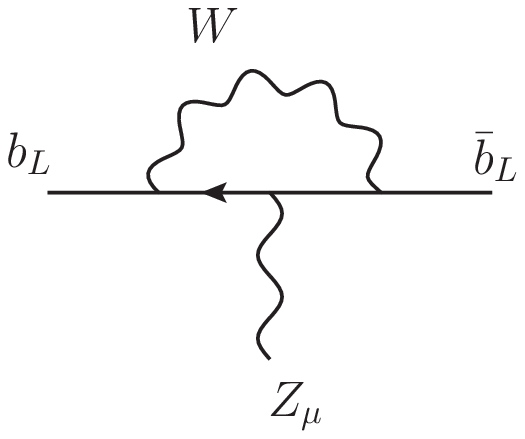}& \includegraphics[scale=0.5]{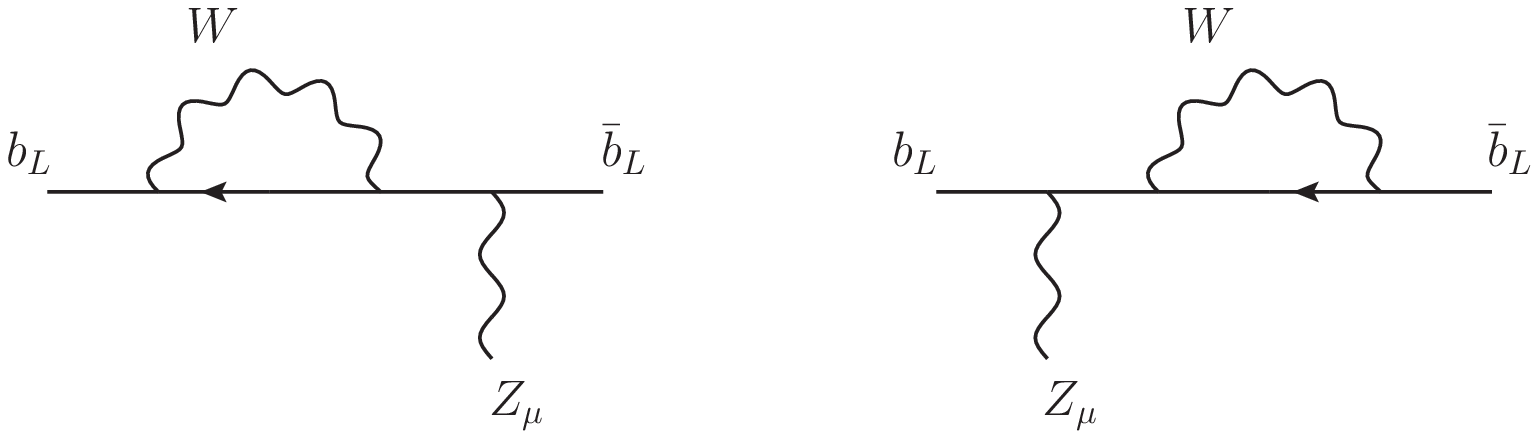}\end{tabular}
	\caption{One-loop diagrams contributing to $Zb_L\bar{b}_L$ in the unitary gauge.
    KK-modes of third generation $Q=2/3$ states and $W^\pm$ zero-mode are
    running in the loop.}
	\label{fig:diagZbbloop}
\end{figure}

We include the one-loop correction to the $Zb_L\bar{b}_L$
coupling as well. The dominant contribution is from KK-fermions
through the diagrams of Fig.~\ref{fig:diagZbbloop}, involving
the SM charged current.
Finally, although such contributions are not present in the
model under study, we report the impact on the fit of the
corrections to the  $S$ and $T$ parameters arising from SM
loops with a pseudo-Goldstone Higgs~\cite{Barbieri:2007bh,
Contino:2010rs}. We refer the reader to appendix~\ref{app:ew}
for further details on both the tree and one-loop calculations.

\subsection{UV Sensitivity of the $S$-parameter}

We start by deriving the 5D degree of divergence of various
one-loop contributions to $S$ using NDA. We match the various
relevant diagrams onto the coefficient $C_S$ of the 5D local
operator, $B_{\mu\nu}W^{\mu\nu a}_LH^{\dagger}\sigma^a H$, that
generates $S$ in the 4D effective action via $S=4\pi v^2
C_S/gg'$. Gauge and fermion contributions to this operator
scale as $C^g_S\propto g_5^4$ and $C^Y_S\propto Y_t^2g_5^2\,$,
respectively. Recalling that the Yukawa coupling has the same
mass dimension as the 5D gauge coupling for a bulk Higgs,
$[Y_t]=[g_5]=-1/2$, power counting yields $C_S^{g,Y}\sim
\Lambda_5^{-1}$, hence a finite contribution, where
$\Lambda_5\equiv N_{\rm KK}\times k$ is the 5D cutoff. Thus $S$
is perfectly calculable at one-loop, and is dominated by the
KK-fermion contribution, provided $Y_t\gg g_5$.

The 5D top Yukawa grows fast in the UV and quickly becomes
non-perturbative. A conservative approach usually requires
$N_{\rm KK}\gtrsim 3$, so that the 5D construction makes sense
as an effective theory; we choose $N_{\rm KK}=3$ in the
following. Assuming in addition that the KK-fermion coupling to
a bulk Higgs is $\mathcal{O}(1)$ for $\beta=0$, NDA yields a
perturbativity upper bound on $Y_t$ of
\beq
Y_t\sqrt{k}\leq 4\pi/\sqrt{N_{\rm KK}}\simeq 7.3\,.
\eeq
\begin{figure}[h!bt]
	\centering
\includegraphics[scale=0.5]{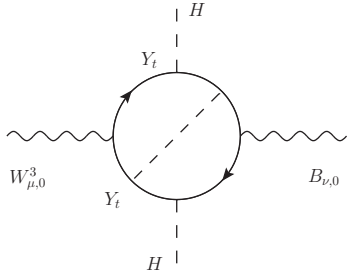}
	\caption{Two-loop diagram relevant for matching onto the
    dimension-six operator generating the $S$ parameter. A similar
    diagram with exchange of weak gauge boson is also present.}
	\label{fig:diagS2loopNDA}
\end{figure}
Higher loops, however, will still be divergent, as they involve
more powers of $Y_t$ and/or $g_5$. This introduces a UV cutoff
sensitivity, even for a bulk Higgs, starting at the two-loop
level. Nonetheless, we argue that the $S$ parameter calculation
is still under control. Indeed, as exemplified by the diagram
shown on Fig.~\ref{fig:diagS2loopNDA}, the two-loop correction
scales like $Y_t^4$ or $Y_t^2 g_5^2$, so its contribution to
$S$ diverges like $\log N_{\rm KK}$. In
Fig.~\ref{fig:diagS2loopNDA} we show only the Higgs as the
internal line. As shown below, this is justified for the sweet
spot parameters, for which $Y_t\sqrt{k}\sim 5$. Hence
contributions from an exchange of KK-gauge bosons will be
subdominant, since  they are proportional to $g^2_5k\sim9$ (see
Appendix~\ref{app:ew}), leading to a $g_5^2/Y_t^2\sim 36\%$
correction NDA then yields
\beq
S_{\rm 2-loop}^{\rm NDA}\simeq \frac{4\pi v^2}{m_{\rm
KK}^2}\frac{N_c}{(16\pi^2)^2}Y_t^4k^2\log N_{\rm KK}\,,
\eeq
where we used the fact that KK-fermion coupling to a bulk Higgs
is $\mathcal{O}(1)$ for $\beta=0$~\cite{Agashe:2008uz}. Thus,
$S_{\rm 2-loop}^{\rm NDA}$ is suppressed by about
$Y_t^2/16\pi^2 \log{N_{\rm KK}} \sim 20\%$ compared to the
one-loop correction. Higher loops will be even more suppressed
since, according to NDA, the expansion parameter is
$Y_t^2\Lambda_5/16\pi^2$, which is smaller than 1 for a
perturbative Yukawa. The $S$ parameter is therefore under
control in our setup.

\subsection{Statistical Analysis}

We first count the parameters of interest in our model.
Imposing the MFV ansatz, the bulk SM flavor symmetries receive
large breaking only from the third generation quark Yukawa
couplings. Then, the  SU(3)$^3_{Q,U,D}$ flavor bulk symmetry is
broken down to an approximate SU(2)$^3_{Q,U,D}$ by the flavon
VEVs, while the lepton flavor group is unbroken. Therefore, the
whole set of effective operators in the SM is determined by 9
free parameters, which we choose to be the fermion bulk masses:
$c_{Q^3}$, $c_t$, $c_b$, $c_{Q^i}$, $c_{U^i}$ and $c_{D^i}$
($i=1,2$, with universal first two generation masses) for the
quark sector, $c_L$ and $c_E$ for the leptons (also taken to be
family universal), and the KK scale, $m_{\rm KK}$. The flavon
VEVs are set by the SM fermion masses.

The global fit analysis proceeds as follows. First of all, a
$\chi^2$-distribution is constructed  by comparing the
experimental measurements to the theoretical predictions of the
model; it is therefore a function of the new physics
parameters: $\chi^2=\chi^2(x)$, where, in our case, $x$
collectively denotes the 9 parameters listed above. The most
probable parameter values, $\bar{x}$, are then identified by
minimizing the total $\chi^2$ function with respect to the
model parameters: $\chi^2(x=\bar{x})\equiv\chi^2_{\rm min}$.
Finally, we bound the parameters $x$ to lie within confidence
level regions around $\bar{x}$, whose size and shape are
dictated by the $\chi^2$ difference, $\Delta\chi^2(x)\equiv
\chi^2(x)-\chi^2_{\rm min}$. The value of $\Delta\chi^2$ is
fixed as a function of the chosen confidence level (CL) and the
number of simultaneously constrained parameters.

\begin{figure}[t]
\centering
\includegraphics[scale=0.5]{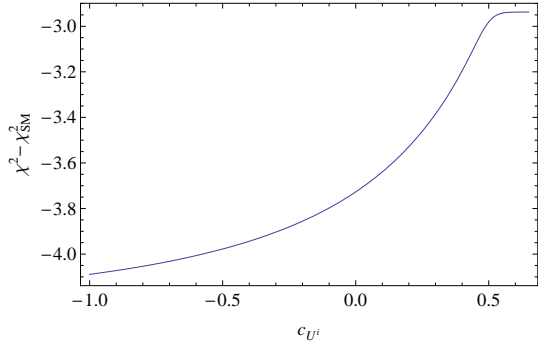}
\caption{$\chi^2 - \chi^2_{SM}$ as a function of $c_{U^i}$, with the rest of
the parameters fixed to the best fit values of Eq.~(\ref{bestfit}). Besides being
relatively insensitive to the localization of $U^i$, the $\chi^2$ distribution
flattens for $c_{U^i}<-0.5$. Here we take $m_{H} = 115~{\rm GeV}$.}
\label{fig:chi2vscUi}
\end{figure}

For this analysis we assume a light Higgs and fix, for
definiteness, its mass to $m_{H} = 115~{\rm GeV}$. We find that
the ``best fit'' parameters in the present scenario are
\beq\label{bestfit}
m_{\rm KK} = 3.5~{\rm TeV}~,~c_t \simeq 0.47~,~ c_b \simeq
0.6~,~ c_{Q^i} \simeq 0.54~, ~ c_L \simeq 0.47~,~ c_e \simeq
0.50~,
\eeq
with a considerably lower sensitivity of the fit to $c_{Q^3}$,
$c_{U^i}$ and $c_{D^i}$ (at the minimum of the $\chi^{2}$, we
find $c_{Q^3} \simeq 0$ and $c_{D^i}\simeq 0.76$). As shown in
Fig.~\ref{fig:chi2vscUi}, there is a preference for $U^i$ to be
composite, although the $\chi^2$ does not depend strongly on
$c_{U^i}$ when $U^i$ is sufficiently IR localized. The values
given in Eq.~(\ref{bestfit}) correspond to $c_{U^i}\simeq
-0.5$, which we will use as a benchmark point. In addition, we
impose the restriction $c_{b} \leq 0.6$ in order to ensure that
the 5D bottom Yukawa coupling is sufficiently small, so that
the $Q = -1/3$ states give a negligible contribution to $\delta
g_{Z\bar{b}_{L}b_{L}}$ (we have not included such
contributions; the required calculation can be extracted
from~\cite{Anastasiou:2009rv}). Note, however, that the fit
prefers a value at the allowed upper limit for $c_{b}$.

The goodness-of-fit of the above model is found to be
$\chi^{2}_{\rm min}/{\rm d.o.f.} = 217.3/223 \approx 0.97$.
This can be compared to the goodness-of-fit of the SM with a
Higgs mass $m_{H} = 90~{\rm GeV}$ (currently the best fit
value): $\chi^{2}_{SM}/{\rm d.o.f.} = 219.9/232 \approx 0.95$.
Thus, the agreement of this particular beyond the SM scenario
is quite comparable to the SM.\footnote{The net decrease in the
total $\chi^{2}$ with respect to the SM can be traced to
$\sigma_{\rm had}$, $R_{e}$, $R_{\mu}$, $A_{e}$, and to a
number of LEP II cross sections. Conversely, we find a worse
fit to the forward-backward asymmetry of the bottom,
$A^{(0,b)}_{FB}$, and to a lesser extent to $R_{b}$. We also
assume in our fit a Higgs mass $m_{H} = 115~{\rm GeV}$, but
this has a negligible impact on the total $\chi^{2}$. For
example, the SM with $m_{H} = 115~{\rm GeV}$ has $\chi^2 =
221.3$, which is only about $1.4$ larger than the value given
above.} We note, however, that we did not fit the SM input
parameters in Eq.~(\ref{bestfit}), but rather fixed them to
their best fit values in the absence of new
physics~\cite{PDG2010}. We proceed now to set CL limits for
models that deviate from Eq.~(\ref{bestfit}).

In our scenario, $\Delta\chi^2$ is a function of 9 new input
parameters, displaying a smooth decoupling limit, with
approximately $\Delta\chi^2\propto m_{\rm KK}^{-2}$. We choose
to present bounds at the $95.4\%$ (2$\sigma$)~CL. We describe
two statistical treatments, of  distinct physical relevance, to
bound the KK scale from EWPTs at a given confidence level (CL).
First of all, we derive a one d.o.f.~bound on $m_{\rm KK}$
only, by marginalizing\,\footnote{Assuming the parameters to be
Gaussian distributed, marginalizing over the bulk masses boils
down to setting them to the values that minimize the $\chi^2$
as a function of the KK scale: {\it i.e.} $c_i=c_i(m_{\rm
KK})$, where the $c_i(m_{\rm KK})$'s satisfy a null gradient
condition $\partial\chi^2/\partial c_i\big|_{m_{\rm KK}}=0$.
(See {\it e.g.}~\cite{NumericalRecipes}.)} over all the bulk
masses and imposing $\Delta \chi^{2} = 4.00$. In addition, we
quote a bound on the KK scale resulting from a simultaneous fit
of the most relevant model parameters. We adopt the following
simple criterion for assessing the relevance of a given input
parameter: if $\partial \log \Delta \chi^{2}/\partial \log
c_{i} > {\cal O}(1)$, then the parameter $c_{i}$ is relevant in
setting the CL. We find that the logarithmic derivatives with
respect to $c_{Q^3}$, $c_{U^i}$ and $c_{D^i}$ are much smaller
than 1 (so we do not count them as d.o.f.~for setting the CL
limits), while the rest of the logarithmic derivatives are
order 1 or larger. Therefore, the second bound on $m_{{\rm
KK}}$ is obtained by assuming $9-3 = 6$ d.o.f., which
translates into $\Delta \chi^{2} = 12.8$ for $95.4\%$ CL.

We stress that these two bounds have a meaning of their own and
contain complementary information.
For the one d.o.f.~analysis, we have that, statistically,
$95.4\%$ of the models show a KK scale larger than the bound,
{\it without any assumptions on all the other parameters}.
Thus, we expect the one d.o.f.~bound on the KK scale to be
rather conservative and of most relevance in terms of LHC
discovery potential. On the other hand, the six d.o.f.~analysis
informs us on the possible correlations among the model
parameters and, in particular, on the existence of less
constrained directions in the parameter space. The presence of
the latter could allow for a lower $m_{\rm KK}$, provided some
other parameters deviate from their best fit values in a
correlated way. As a result, however, we expect such a KK scale
to be statistically unlikely, although we have not tried to
quantify this statement. Nevertheless, we think that the
existence of such points in the parameter space are worth
mentioning, for such correlation may be theoretically motivated
and/or future experimental analyses may become sensitive to
additional parameters, on top of the KK scale.

\subsection{EWPT Global Fit Results}

We report in this section the sweet spots found for the one and
six d.o.f.~statistical analyses defined above. These are done
for both our flavor-triviality model, as well as for a (slight)
variant of the conventional anarchic model.

\subsubsection{Sweet spot in the flavor-triviality model}
Assuming the hierarchical Yukawa ansatz, we find from the
global fit a bound on the KK scale of $m_{\rm KK}> 2.1 \textrm{
TeV}$ $(95.4\%~{\rm CL})$ for the one d.o.f.~analysis
(i.e.~$\chi^{2} - \chi^{2}_{\rm min} = 4.00$). The
corresponding ``sweet spot'' values for the bulk masses are
\beq\label{1dsweet}
&&c_{Q^3}\simeq 0.05 \,, \ c_t\simeq 0.47 \,, \ c_{Q^i} \simeq
0.51 \,, \ c_L \simeq 0.48 \,, \ c_e \simeq 0.50 \,,\nonumber\\
&& c_b\simeq 0.6\,, \ c_{U^i}\simeq -0.5\,, \ c_{D^i}\simeq
0.77~.
\eeq
If instead the SM is taken as the ``best fit point'' (i.e.
$\chi^{2} - \chi^{2}_{\rm SM} = 4.00$), the resulting bound is
improved to 1.8~TeV (with some changes in the bulk masses).
Performing a six d.o.f.~analysis at 95.4\%~CL (i.e. $\chi^{2} -
\chi^{2}_{\rm min} =12.8$) yields $m_{\rm KK}> 1.7 \textrm{
TeV}$, with
\beq\label{6dsweet}
&&c_{Q^3}\simeq 0.02 \,, \ c_t\simeq 0.48 \,, \ c_{Q^i} \simeq
0.50 \,, \ c_L \simeq 0.48 \,, \ c_e \simeq 0.50 \,,\nonumber\\
&& c_b\simeq 0.6\,, \ c_{U^i}\simeq -0.18\,, \ c_{D^i}\simeq
0.77~.
\eeq

\begin{figure}[h!]
\centering
\includegraphics[scale=0.5]{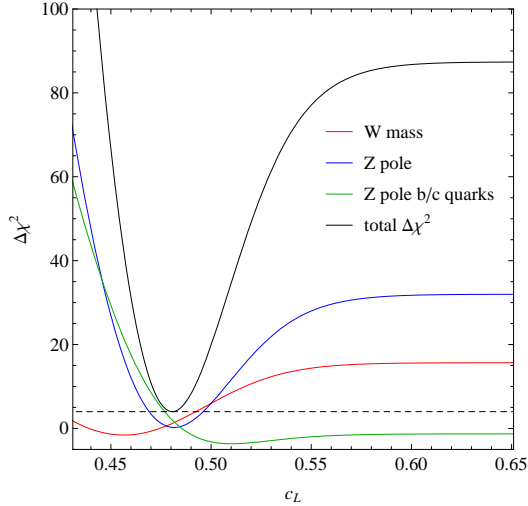}
\caption{Most important contributions to $\Delta\chi^2$
as a function of $c_L$, with $m_{\rm KK}$ and other bulk masses set to the sweet spot values of Eq.~(\ref{1dsweet}). $Z$ pole
observables (blue) include the total $Z$ width, $e^+e^-$ hadronic cross
section and other leptonic observables, while the heavy quark observables (green) include $R_{b,c}$, $A_{b,c}$ and $A_{b,c}^{FB}$.}
\label{fig:chi2}
\end{figure}

We illustrate in Fig.~\ref{fig:chi2}, as a function of the LH
lepton localization parameter, $c_{L}$, the $\Delta\chi^2$
contributions which are most sensitive to this parameter; they
are the Z pole observables (including b and c quark
observables) and the W mass measurements. This shows that a low
KK scale is achieved for relatively flat light fermions,
$c_L\simeq 0.48$, as expected from cancellation of an effective
$S$ parameter (see also Eq.~\eqref{sft} below). We also report
in Table~\ref{tab:chi2pieces} the contributions to the $\chi^2$
and $\Delta\chi^2$ for the one d.o.f.~sweet spot of
Eq.~(\ref{1dsweet}).
\begin{table}[htbp]
	\centering
	\begin{tabular}{|c|c|c|c|}\hline
		& $\ \chi^2-\chi^2_{\rm SM}\ $ & $\ \chi^2_{\rm min}-\chi^2_{\rm SM}\ $ & $\ \Delta\chi^2 \ $ \\ \hline\hline
W mass &1.37 &0.12& 1.25 \\
Z line shape \& lepton $A_{FB}$ & $-$2.47& $-$2.74& 0.27\\
Z pole b\&c quarks & 5.30& 3.44& 1.86 \\
$s_W^2$ hadronic charge asymmetry &0.20&0.23&$-$0.03 \\
Leptonic polarization asymmetries &$-$1.95& $-$2.18& 0.23 \\
Deep inelastic scattering & $-$0.13& $-$0.12& $-0.01$ \\
Atomic parity violation & 3.23&0.11 & 3.12 \\
 LEP2 hadronic cross-section &$-$1.97&$-$0.97& $-$1.00\\
 LEP2 muon pair &$<10^{-2}$& 0.03 & $-$0.03\\
 LEP2 tau pair &$-$0.04 & $-$0.03& $-$0.01\\
 OPAL electron pair & $-$0.02& $-$0.02 & $<10^{-2}$ \\
 L3 W pair &$-$0.17&$-$0.11& $-$0.06\\
 Z pole s quark &0.07& 0.09&$-$0.02\\
 LEP2 ee$\to$bb &$-$3.22&$-$1.75 & $-$1.47\\
 LEP2 ee$\to$cc &$-$0.18&$-$0.08& $-$0.10\\ \hline \hline
 Total & 0.02 & $-$3.98 & 4 \\ \hline
\end{tabular}
\caption{Contributions to the $\chi^2$ and $\Delta\chi^2$ for
the sweet spot of Eq.~(\ref{1dsweet}); see~\cite{Han:2004az}.}
\label{tab:chi2pieces}
\end{table}
\subsubsection{Bounds on the semi-anarchic model}
For the sake of comparing our setup to known anarchic models,
and better assessing the benefits of the flavor triviality
scenario, we report the EW global results for a
``semi-anarchic" model defined as follows. We set the first two
quark generations and all the leptons to be elementary,
$c_{Q^i}=c_{U^i}=c_{D^i}=c_L=c_E=0.65$, thus allowing the
corresponding 5D Yukawa couplings to be all of the same order
(the fit is completely insensitive to the precise value of the
$c$'s, or to the fact that these are all the same, as long as
they are UV localized). However, we require $c_b < 0.6$, so
that the 5D bottom Yukawa coupling is suppressed. This
restriction ensures that our loop contribution to $\delta
g_{Z\bar{b}_{L}b_{L}}$ is reliable, as mentioned above. The
results of relaxing this assumption, so that full anarchy can
be achieved, will be presented elsewhere.  The parameters
determined in the fit are $m_{\rm KK}$, $c_{Q^3}$, $c_{t}$ and
$c_{b}$. Unlike what was found in the flavor-triviality model,
under the anarchy assumption the minimum $\chi^{2}$ is obtained
when $m_{{\rm KK}} \to \infty$, hence it is the same as in the
SM. The goodness-of-fit in this case is $\chi^{2}_{\rm SM}/{\rm
d.o.f.} = 221.3/228 \approx 0.97$.

We then find a one d.o.f.~bound ($\chi^{2} - \chi^{2}_{\rm SM}=
4.00$) on the KK scale of $m_{\rm KK}> 4.6 \textrm{ TeV}$
$(95.4\%~{\rm CL})$ with the following sweet spot values
\beq\label{1danarchy}
c_{Q^3}\simeq 0.11 \,, \ c_t\simeq 0.49 \,, \ c_b\simeq 0.6 \,.
\eeq
For completeness, we also report that comparing the
semi-anarchic scenario to the best fit point of
Eq.~(\ref{bestfit}), the bound is raised to $m_{\rm KK} \gtrsim
7$~TeV ($\chi^{2} - \chi^{2}_{\rm min}= 4.00$).

In order to set a limit on $m_{{\rm KK}}$ by simultaneously
fitting all the parameters, we note that $m_{{\rm KK}}$ and
$c_{t}$ are unequivocally relevant parameters (as defined in
the previous subsection), while the logarithmic derivative of
$\Delta \chi^{2}$ with respect to $c_{Q^3}$ is much smaller
than 1, and the one corresponding to $c_{b}$ is of order 1. We
therefore perform a $4-1 = 3$ d.o.f.~analysis, corresponding to
$\Delta \chi^{2} = 8.02$ for 95.4\%~CL. This yields $m_{\rm
KK}> 3.9~\textrm{TeV}$ with
\beq\label{2danarchy}
c_t\simeq 0.49 \,, \ c_b \simeq 0.6 \,,
\eeq
where we fixed $c_{Q^3}=0.10$.

We end this subsection by emphasizing that in this work we
explore the possibility that the fermions span $SU(2)_{L}
\times SU(2)_{R}$ representations. However, the loop-level
contributions to $S$, $T$ and $\delta g_{Z\bar{b}_{L}b_{L}}$
can be rather dependent on this assumption. For instance, when
the third generation fermions are assigned to $SO(5)$
representations, \`a la gauge-Higgs unification, one finds that
the one-loop S-parameter can be significantly smaller than for
the $SU(2)_{L}\times SU(2)_{R}$ representations (this happens,
{\it e.g.}, in the scenario of Ref.~\cite{Carena:2007ua}). This
can affect the bounds for the anarchic scenario, which are
controlled by the oblique parameters (plus $\delta
g_{Z\bar{b}_{L}b_{L}}$). As an example, in the scenario
discussed in \cite{Carena:2007ua}, where the corresponding
bound was found to be $m_{\rm KK}> 3.4~{\rm TeV}$, an updated
3-parameter fit to the EW data leads to $m_{\rm KK}> 3~{\rm
TeV}$ (both cases are compared to the SM as the best fit). This
slight improvement is mainly due to the use of the most recent
SM fit, that has moved in a favorable direction for these
scenarios. Regarding the flavor triviality case, we expect that
the representation assignments for the third family are less
crucial for the sweet spot, since the difference in the
loop-level S-parameter can be compensated to some extent by an
effective tree-level contribution (with a slight readjustment
of parameters). What is more important for the sweet spot is
the custodial protection associated with the lepton
representations, as emphasized in the introduction. With these
caveats, we explore the degree of tuning involved in the next
subsection.

\subsection{Fine-Tuning Estimates}

The bulk RZ model displays a sweet spot of bulk masses where
the KK scale is significantly lower than for the most
optimistic anarchic cases. Such a lower KK scale would, in
principle, reduce the fine-tuning associated with the Higgs
mass in warped models. However, this result is a mere
consequence of the approximate flatness of the SM light fermion
wave functions. We therefore expect a large sensitivity of the
KK scale under corrections to the sweet spot values of the bulk
masses, and a potentially larger new source of fine-tuning. We
show that, even in the anarchic case, a sensitivity of this
sort is actually also present; we shall estimate its size as
well.

Because of the flavor symmetries, the only UV sensitive
contributions are expected to be related to gauge interactions,
which distinguish between different fermion representations.
This would raise a legitimate question regarding the sweet
spot: how come fields related to different SM representations
are located near each other, in particular around $c_x\sim
0.5$? We do not have a sharp answer to this question. However,
one could imagine embedding the above theory into some form of
unification model  (for an SO(10) grand unified theory (GUT)
see for example~\cite{Randall:2001gb,Pomarol:2000hp,
Randall:2001gc, Agashe:2004ci, Agashe:2004bm, Agashe:2005vg,
Agashe:2009ja}), which would explain why the couplings are
related to each other.\footnote{This would require various
fermions to come from the same GUT multiplet, which would
require a non-conventional approach to ensure proton longevity.
Alternatively, one could impose a discrete symmetry that would
correspond to invariance with respect to interchanging the
different GUT multiplets.} Finite radiative corrections to
these quantities are proportional to the bulk masses
themselves~\cite{Cheng:2002iz}. The radiative corrections to
the bulk masses, which split the universal part of the
fermions' wave-functions, $c_i$, will be finite and suppressed
by a loop factor of order $g^2_5 k /16 \pi^2$ ($g^2_5 k\sim9$).
Therefore, a mass splitting of a few percent is expected. It is
interesting that within the RS framework the radiative
correction to the masses seems to vanish for flat fermions.
However, we shall not pursue this possibility, as it goes
beyond the scope of this project.

We now estimate the fine-tuning in the flavor triviality and
semi-anarchic models. The fine-tuning is composed of two
ingredients, namely the sensitivity of the weak scale to the KK
scale, $FT_{m_{\rm KK}}$, and the sensitivity of the KK scale
itself to the bulk masses,  $FT_c$, through the global EW fit.
Strictly speaking, the former is under control only if the
Higgs is a pseudo-Goldstone boson ({\it i.e.}, if its mass is
finite). Nonetheless, we believe a pragmatic
approach~\footnote{In the present model the Higgs mass
parameter is quadratically sensitive to the cutoff scale,
rather than to the KK scale. Our intention here is to very
roughly estimate the fine-tuning of the EW scale in PGB
extensions that also incorporate the ingredients discussed in
this paper. One should remember, however, that such extensions
can contain additional correlations that may not allow a KK
scale as low as we have found above, or may contain additional
sources of fine-tuning (see e.g.~\cite{Panico:2008bx}).
Nevertheless, we believe that the new ingredients highlighted
here should help in relaxing the bound on $m_{\rm KK}$, and
hence associated the fine-tuning in such models.} consists in
estimating this fine-tuning as $FT_{m_{\rm KK}}\simeq
(v/f_\pi)^2$ where $f^{-1}_\pi \approx R'$~(see {\it
e.g.}~\cite{Agashe:2004rs}). The specific definition is not
crucial (and different studies differ in their order 1
coefficients in any case~\cite{Agashe:2004rs, Csaki:2008zd,
Panico:2008bx}), but it does enable us to assess the difference
in success between the anarchic and flavor triviality cases.

Regarding $FT_c$, one conventional procedure is to relate it to
the logarithmic derivative at the sweet spot, $ S_{c} \equiv
\max_i\left|\partial \log m_{\rm KK}/\partial \log
c_i\right|$~\cite{Barbieri:1987fn,Anderson:1994dz}, and to
interpret its inverse as a measure of the fine-tuning involved.
However, since the sweet spot naturally resides in a local
minimum of the parameter space, this derivative exactly
vanishes. Instead, we find it convenient to use the (one-sided)
finite difference analog, which gives a measure of the
sensitivity of $m_{\rm KK}$ to $c_i$ in a vicinity of the sweet
spot. Since these one-sided finite derivatives can be different
on both sides of the best fit point, especially for parameters
which control the size of the top Yukawa coupling ({\it e.g.}
$c_t$), we use an average of the two,
\be \label{discrete_log_derivative}
S_{c_i} \equiv \frac{1}{2}\, \left(\left| \frac{c_i} {\Delta
c_i} \, \frac{\Delta m_{\rm KK}^+}{m_{\rm KK}} \right| + \left|
\frac{c_i} {\Delta c_i}\, \frac{\Delta m_{\rm KK}^-}{m_{\rm
KK}} \right|\right) ,
\ee
where we choose $\Delta c_i=0.03\,$, as motivated by the
typical size of the radiative corrections. Here $\Delta m_{\rm
KK}^\pm=m_{\rm KK}(c_i \pm \Delta c_i)-m_{\rm KK}(c_i)$ is the
change of the KK scale for a given $\Delta c_i$, with the other
bulk masses fixed, that is necessary to keep $\Delta \chi^2$
fixed (so as to keep the success of the EW global fit at the
same level). The final sensitivity should correspond to the
largest value obtained by repeating the procedure for all the
parameters in the model, $S_c={\rm max}_i S_{c_i}$.

Using the definitions above, we find for the flavor triviality
(one, six) d.o.f.~sweet spots the following measures of
fine-tuning:
\beq \label{FTsweet}
FT_{m_{\rm KK}}\simeq (8.4,13)\%~; \hspace{5.2cm}
\nonumber \\
S^{-1}_{c_t}\simeq (5.3,\,10)\%~;~~ S^{-1}_{c_L}\simeq
(6.2,\,8.9)\%~;~~ S^{-1}_{c_E}\simeq (8.1,\,9.3)\%~;~~
S^{-1}_{c_{Q^i}}\simeq  (46,\,42)\%~,
\eeq
while there is essentially no sensitivity to the rest of the
input parameters. If the SM is assumed to constitute the best
fit, these numbers slightly improve,
\beq \label{FTsweet_SM}
FT_{m_{\rm KK}}\simeq (11,14)\%~; \hspace{5.2cm}
\nonumber \\
S^{-1}_{c_t}\simeq (8.1,\,10)\%~;~~ S^{-1}_{c_L}\simeq
(8.2,\,9.1)\%~;~~ S^{-1}_{c_E}\simeq (9.2,\,9.4)\%~;~~
S^{-1}_{c_{Q^i}}\simeq  (44,\,42)\%~,
\eeq
In contrast, for the semi-anarchic model with fermions in
$SU(2)_L \times SU(2)_R$ representations, we find
\beq \label{FTanarchy}
\textrm{SM:}&& \quad FT_{m_{\rm KK}}\simeq (1.7,\,2.3)\%~;~~
S^{-1}_{c_t}\simeq (22,\,20)\%~, \nonumber \\ \textrm{Best
fit:}&& \quad FT_{m_{\rm KK}}\simeq (0.7,\,1.7)\%~;~~
S^{-1}_{c_t}\simeq (-,\,22)\%~,
\eeq
for (one, three) d.o.f., respectively.\footnote{Due to the high
KK scale observed in the semi-anarchic model when compared to
the best fit point, see below Eq. \eqref{1danarchy}, the 1
d.o.f.\ requires us to extrapolate the results, hence the
sensitivity in that case could not be computed. However, from
the three d.o.f. case we see that the sensitivity is roughly
the same as when the bound was compared to the $\chi^2_{\rm
SM}$.} We then see that indeed the fine-tuning of the weak
scale is improved in our model. On the other hand, the
sensitivity to the bulk masses is greater. We regard the
sensitivity exhibited by $S_c^{-1}$ as an indication of
fine-tuning that, together with $FT_{m_{\rm KK}}$, determines
the overall fine-tuning of the model.

Finally, it is interesting to analytically examine the
sensitivity of the vertex corrections [see
Eqs.~(\ref{eqn:O6hf},\ref{as})] to the localization of the
leptons in our model, which is the main source for
$S_{c_{L,E}}^{-1}$, as given above. The parametric dependence
on $c_{L,E}$ and $m_{\rm KK}$ of the corresponding operators is
\be
a_{hF}^{t,s} \propto \frac{(c_{L,E}-1/2+k)}{m_{\rm KK}^2} \,,
\ee
where $c_{L,E}-1/2$ originated from the $\{++\}$ gauge
KK-states~\cite{Agashe:2003zs} and $k \simeq 0.06$ effectively
parametrizes the contribution of the $\{-+\}$ gauge KK-states.
Using this expression, we can approximately evaluate
$S_{c_{L,E}}^{-1}$ as
\be \label{sft}
S_{c_{L,E}}^{-1}= \left( \frac{ \partial \log m_{\rm KK}}{
\partial \log c_{L,E}} \right)^{-1} \simeq \frac{2(c_{L,E}-1/2+k)}{c_{L,E}}~,
\ee
which for $c_L=0.48$ gives $S_{c_L}^{-1}
\simeq 0.16\,$. The remaining sensitivity above comes from the
rest of the observables.

\section{Flavor physics} \label{sec:flavor}

Our setup is a variation of the anarchic 5DMFV
model~\cite{Csaki:2009wc,Fitzpatrick:2007sa}, where the shined
Yukawas are of hierarchical structure as
in~\cite{Rattazzi:2000hs}, but the SM quarks propagate in the
bulk. Therefore, the following relation between the bulk masses
and the 5D Yukawa matrices is obtained:
\be \label{bulk_masses_short}
\begin{split}
C_Q&=a_Q\cdot \mathbf{1}_3+b_U^Q \,Y_U Y_U^{\dagger}+b_D^Q
\,Y_D Y_D^{\dagger}+\dots \,,\\ C_{U,D}&=a_{U,D}\cdot
\mathbf{1}_3+b_{U,D}\, Y_{U,D}^{\dagger}Y_{U,D}+\dots \,,
\end{split}
\ee
where the dots stand for contributions from higher powers of
the Yukawa flavons. Recall also that, in the absence of mixing,
the masses in terms of the Yukawas are given by
\begin{equation} \label{mass0}
m_{U,D}\simeq\alpha_{U,D} \frac{v}{\sqrt2} \,F_Q Y_{U,D} F_{U,D}
\,r^H_{00}(\beta,c_Q,c_{U,D})+\dots \,,
\end{equation}
where $F_X$ are matrices with eigenvalues $f_{x^i}$
representing the IR projection of the quark zero-mode profiles,
given by $f_{x^i}^2=(1-2c_{x^i})/( 1-\epsilon^{
1-2c_{x^i}})\,$, $c_{x^i}$ are the eigenvalues of the $C_x$
matrices, $\epsilon=\exp[-\xi]$, $\xi=\log[M_{\rm
\overline{Pl}}/{\rm TeV}]$, $M_{\rm \overline{Pl}}$ is the
reduced Planck mass and $r^H_{00}(\beta,c_L,c_R) \approx
\frac{\sqrt{2(1+\beta)}}{2+\beta-c_L-c_R}$ is the overlap
correction for a bulk Higgs~\cite{Gedalia:2009ws}
($r^H_{00}(\beta,c_L,c_R)=1$ for a brane-localized Higgs). The
$\alpha_{U,D}$ coefficients are distinct from those in the
expansion of Eq.~(\ref{bulk_masses_short}) (in our subsequent
discussion, only the combinations $\alpha_{U,D} Y_{U,D}$
appear). For simplicity we show in Eq.~\eqref{mass0} only the
part related to the zero-mode couplings and the leading term in
terms of the Yukawa flavon fields. In practice, the third
generation masses are somewhat modified due to the fact that
the mass eigenstates are affected by mixing with the
KK-fermions, hence this is taken into account in our
quantitative analysis. NDA suggests that in the most generic
models $b_{U,D}^Q$, $b_{U,D}$ and $\alpha_{U,D}$ are all of
order 1 in appropriate units of the
curvature~\cite{Csaki:2009wc}. However, we point out that
$\alpha_{U,D}$ carry different $U(1)_{Y_{U,D},\bar Q,U,D,H}$
charges (which can be thought of as generalized Peccei-Quinn
symmetries), and therefore a hierarchy between $\alpha_U$ and
$\alpha_D$, and between $\alpha_i$ and $b_i^Q,b_i$ is natural,
and can be obtained in specific models. For instance, in models
of gauge-Higgs unification, $\alpha_i$ can be indirectly
suppressed due to gauge interactions.

An immediate consequence of the MFV framework is that bounds
from flavor violation in the first two generations become much
weaker. This follows from an inherent suppression of
right-handed currents, which require light mass
insertions~\cite{Kagan:2009bn,D'Ambrosio:2002ex}. Thus, the
bound from $\epsilon_K\,$, which is rather strong in the
anarchic case~\cite{Bona:2007vi,Davidson:2007si, Csaki:2008zd,
Blanke:2008zb,Bauer:2009cf}, is irrelevant
here~\cite{Kagan:2009bn}, since the right-handed current is
suppressed by $r_Q^4\, m_s m_d/m_b^2$ ($r_Q\sim 2-3$, see
Eq.~\eqref{rq} in Appendix~\ref{app:flavor}) compared to the
left-handed current.

As a result of the large top mass, we actually expect higher
powers of the up Yukawa to be important, and these would shift
the eigenvalues of the bulk masses~\cite{Kagan:2009bn,
Gedalia:2010rj}. The impact of top Yukawa resummation is
subtle, but can be observed in flavor violation involving
left-handed currents in the first two generations. This
applies, in particular, to CP violation in the D
system~\cite{Gedalia:2009kh} (effects of order $m_c^2/m_t^2$
are present in the kaon system, but are much harder to
observe~\cite{Kagan:2009bn,Buras:2010pz}). If the bottom Yukawa
is large as well, then in the presence of flavor diagonal
phases, order 1 CP violating contributions are expected in
$B_{d,s}$ mixing~\cite{Kagan:2009bn,Ellis:2007kb,
Colangelo:2008qp,Mercolli:2009ns,Paradisi:2009ey}. An easy way
to see this is to take the two generation limit, where the SM
Lagrangian is manifestly CP conserving. In this case, higher
dimensional operators can contain a CP violating combination of
the Yukawa matrices, proportional to the covariant flavor
direction, $\hat J$~\cite{Gedalia:2010zs,Gedalia:2010mf}
\begin{equation}
\hat J\propto \left[Y_D Y_D^\dagger,Y_U Y_U^\dagger\right]\,.
\end{equation}
This induces CP violation in both up and down sectors, even in
the two generation case.

We shall focus on two scenarios. The first is when the bulk
parameters give a small 5D bottom Yukawa. Generically, the
phenomenology of this model is rather simple, and the
contributions to various flavor changing processes are highly
suppressed. We then slightly deform this sweet spot solution,
and show how the model approaches the large bottom 5D Yukawa
limit, which yields a richer flavor structure. In particular,
we demonstrate how one can generate sizable new CP violating
contributions in $B_d$ and $B_s$ mixing, and identify a natural
region of the parameters where the latter dominates, as favored
by the recent D$\slashed{0}$~data~\cite{Abazov:2010hv,
Abazov:2008fj} and permitted by the CDF one~\cite{CDF:2010}
(see~\cite{Ligeti:2010ia,Dobrescu:2010rh,Buras:2010mh,
Jung:2010ik, Eberhardt:2010bm,Dighe:2010nj, Chen:2010wv,
Bauer:2010dg, Deshpande:2010hy, Batell:2010qw, Kurachi:2010fa,
Chen:2010aq,Parry:2010ce} for related work
and~\cite{Randall:1998te} for a much earlier study about lepton
asymmetry in the $B$ system).

In the following we employ only a 1 d.o.f.\ analysis of the
EWPT bounds, for simplicity. We also mainly focus on a
comparison with the best fit point (which is now required to
comply with flavor constraints).

\subsection{Small 5D Bottom Yukawa} \label{sec:small_yb}

We first analyze the flavor structure of the theory with a
small bottom Yukawa. For concreteness we give an example of
such a point\footnote{In the context of flavor physics, it is
more convenient to employ the notations $c_{U^3}$ and $c_{D^3}$
instead of $c_t$ and $c_b$ used above. We thus adopt this
change in this section.},
\begin{eqnarray} \label{bulk_masses_ss}
C_Q &=& (0.497, 0.497, 0.348)\,, \ \ \  C_U = (-0.5, -0.5, 0.482)\,, \ \ \
C_D = (0.56, 0.56, 0.55)\,,\nonumber \\
\alpha_U Y_U &=& (3.6 \times 10^{-5}, 0.017, 6.2)\,, \ \ \ \alpha_D Y_D = (0.0013, 0.024, 0.36)\,.
\end{eqnarray}
The reader should bear in mind though, that as long as the
bottom Yukawa is small, the gross features of the model near
the sweet spot remain unchanged. The resulting bound from EWPTs
is 2.4~TeV.

In the limit of small $Y_D$, the bulk masses can be expanded in
powers of $Y_U$ only. This is manifest in the choice of bulk
masses in Eq.~\eqref{bulk_masses_ss}, where $C_D$ is almost
completely diagonal, and in $C_{Q,U}$ only the third eigenvalue
is shifted away from the first two. Since this applies to the
$F_X$'s as well, we have according to Eq.~\eqref{mass0}
$\left[m_U,Y_U \right]=0$, {\it i.e.}, $m_U$ and $Y_U$ can be
simultaneously diagonalized. Our model thus contains a built-in
up-type flavor alignment, hence FCNCs are only present in the
down sector. Moreover, flavor violation in the down sector is
proportional to elements of the Cabibbo-Kobayashi-Maskawa (CKM)
matrix $\vckm\,$, and right-handed currents are significantly
suppressed, as anticipated since our current setup belongs to
the MFV framework with a small bottom Yukawa.

It is crucial to emphasize that when expanding the bulk masses
as functions of the Yukawa matrices, higher order terms in $Y_U$ are
important, and may give rise to a significant effect, as shown
below. Therefore, we write
\be \label{bulk_masses_long}
\begin{split}
C_Q&=a_Q\cdot \mathbf{1}_3+b_U^Q \,Y_U Y_U^{\dagger}+b_D^Q
\,Y_D Y_D^{\dagger}+d_{UU}^Q Y_U \left(Y_U Y_U^{\dagger}\right)
Y_U^{\dagger} +d_{DU}^Q Y_D \left(Y_U Y_U^{\dagger}\right)
Y_D^\dagger+ \dots \,, \\ C_U&=a_U\cdot \mathbf{1}_3+b_U\,
Y_U^{\dagger}Y_U+ d_{UU} Y_U^\dagger \left(Y_U
Y_U^{\dagger}\right) Y_U + \dots \,, \\ C_D&=a_D\cdot
\mathbf{1}_3+b_D\, Y_D^{\dagger}Y_D+ d_{DU} Y_D^\dagger
\left(Y_U Y_U^{\dagger}\right) Y_D + \dots \,,
\end{split}
\ee
where some of these terms are actually small in the case of
small bottom Yukawa.

The most severe constraints are from the $B_{d,s}$ systems, in
the form of a $\Delta B=2$ contribution to the mixing
amplitude. These are generated in RS via a tree-level KK-gluon
exchange, formulated in terms of two of the standard four-quark
operators,
\begin{equation} \label{4q_operators}
\begin{split}
Q_1&= \bar q^\alpha_{jL} \gamma_\mu q^\alpha_{iL}
\bar q^\beta_{jL} \gamma_\mu q^\beta_{iL} \,, \\
Q_4&= \bar q^\alpha_{jR} q^\alpha_{iL} \bar q^\beta_{jL}
q^\beta_{iR} \,,
\end{split}
\end{equation}
where $\alpha,\beta$ are color indices and $i,j$ are flavor
indices. New physics in the $B_{d,s}$ mixing amplitudes can be
described by four real parameters,
\be
M_{12}^{d,s}= \left( M_{12}^{d,s} \right)^{\rm SM} \left(
1+h_{d,s} e^{2i \sigma_{d,s}} \right) \,,
\ee
where $M_{12}$ is the dispersive part of the amplitude. We
shall use the notation $h_{d,s}^{1,4}\,$, where the superscript
denotes the contributing operator.

In order to evaluate the flavor-violating contribution to
$B_d$, we need to rotate the diagonal coupling of two quarks
with the KK-gluon to the mass basis. Since the mass basis is
aligned with $Y_U$, this introduces CKM factors (plus
subleading corrections for large bottom Yukawa) in the case of
left-handed quarks, and a factor related to the difference of
overlaps of the $b$ and $d$ quarks with the KK-gluon. This
calculation is performed in detail in Appendix~\ref{app:flavor}
(see Eq.~\eqref{fq_fd_13}). The Wilson coefficient for $Q_1$ is
then given by
\be \label{kkgluon_c1}
C_1 \approx \frac{g_{s*}^2}{6 m_{\rm KK}^2} \left(V_{tb}
V_{td}^* \right)^2 \left[ f_{Q^3}^2 r_{00}^g(c_{Q^3})
-f_{Q^1}^2 r_{00}^g(c_{Q^1}) \right]^2 \,.
\ee
Here $g_{s*}$ is the dimensionless 5D coupling of the gluon
($g_{s*}=3$ with one-loop matching), $r_{00}^g(c) \approx
\frac{\sqrt2}{J_1(x_1)} \, \frac{0.7}{6-4c} (1+e^{c/2})$ is the
overlap correction for two zero-mode quarks with the
KK-gluon~\cite{Gedalia:2009ws,Csaki:2008zd,Csaki:2009bb}, with
$x_1 \cong 2.4$ as the first root of the Bessel function
$J_0(x_1)=0$, and $\left(V_{tb} V_{td}^* \right)^2 \approx
\left[\vckm_{tb} \left(\vckm_{td}\right)^* \right]^2
\left(1+rY_b^2 e^{i2\theta_d}\right)\,$, with $\theta_d$ an
arbitrary phase and $r$ a proportionality coefficient (in the
current case we neglect this correction, which is formally of
order $Y_b^2$). A similar formula applies for $B_s\,$
(replacing $d \to s$ and $1 \to 2$).

For a right-handed coupling, which is a part of the $Q_4$
contribution, the rotation is more involved, and introduces
some additional factors (see Appendix~\ref{app:flavor}). The
resulting Wilson coefficient is
\be \label{kkgluon_c4}
\begin{split}
C_4 ~\approx~ & \frac{g_{s*}^2}{m_{\rm KK}^2} \left(V_{tb}
V_{td}^* \right)^2 \frac{m_d}{m_b}\, \left[
\left(\frac{f_{Q^3}\, r_{00}^H(\beta,c_{Q^3},c_{D^3})
}{f_{Q^1}\, r_{00}^H(\beta,c_{Q^1},c_{D^3})} \right)^2
-1\right]  \\ &
\times
\left[f_{Q^3}^2 r_{00}^g(c_{Q^3}) -f_{Q^1}^2
r_{00}^g(c_{Q^1}) \right]
\left[ f_{D^3}^2
r_{00}^g(c_{D^3}) -f_{D^1}^2 r_{00}^g(c_{D^1}) \right] \,.
\end{split}
\ee

In order to derive a bound on the KK scale, we allow for
$h_d^{1,4}$ to be as large as 0.5 (since the NP contributions
do not carry additional CP phases)~\cite{Ligeti:2010ia}. We
include running and mixing effects at 2 TeV, as described
in~\cite{Bona:2007vi} and refs.~therein. The bound resulting
from $Q_1$ is
\be \label{kkgluon_bd_c1}
\left( \frac{m_{\rm KK}}{2 \rm \,TeV}\right)\gtrsim 3.7
\left(\delta f^2_{Q^{31}}\right) \approx 2.3\,
\left(\frac{1-2.1\, c_{Q^3}}{1-\frac23\, c_{Q^3}} \right) \,,
\ee
where we defined
\be
\left(\delta f^2_{Q^{ij}}\right) \equiv f_{Q^i}^2
r_{00}^g(c_{Q^i}) -f_{Q^j}^2 r_{00}^g(c_{Q^j}) \,,
\ee
and used $c_{Q^1}=0.497$ from Eq.~\eqref{bulk_masses_ss} and
\be
f_x^2 r_{00}^g(c_x) \approx \frac{1-2\,c_x}{1.5-c_x} \,,
\ee
which is a good approximation for $0<c_x<0.47\,$. Note that for
$c_{Q^3}=0.348$ we have $m_{\rm KK} \gtrsim 1.9$~TeV,
consistent with EWPT.\footnote{This is weaker than
in~\cite{Isidori:2010kg}, which was ultra-conservative.}
Similarly, the bound from $Q_4$ is
\be
\left( \frac{m_{\rm KK}}{2 \rm \,TeV}\right) \gtrsim 23\,
\sqrt{ \frac{m_d}{m_b} \left[ \left(\frac{f_{Q^3}\,
r_{00}^H(\beta,c_{Q^3},c_{D^3}) }{f_{Q^1}\,
r_{00}^H(\beta,c_{Q^1},c_{D^3})} \right)^2 -1\right]
\left(\delta f^2_{Q^{31}}\right) \left(\delta
f^2_{D^{31}}\right)}~.
\ee
The actual constraint is much weaker than
Eq.~\eqref{kkgluon_bd_c1} because of the $m_d/m_b$ suppression
and the approximate degeneracy of the $f_D$'s. It is
instructive to see the relation between the contributions of
$Q_4$ and $Q_1$ to $B_d$ mixing:
\be \label{c1c4_bd}
\frac{C_4}{C_1} \Bigg|_{\textrm{2 TeV}} \approx 40\,
\frac{m_d}{m_b} \, \frac{\left(\delta f^2_{D^{31}}
\right)}{\left(\delta f^2_{Q^{31}}\right)} \left[
\left(\frac{f_{Q^3}\, r_{00}^H(\beta,c_{Q^3},c_{D^3})
}{f_{Q^1}\, r_{00}^H(\beta,c_{Q^1},c_{D^3})} \right)^2
-1\right] \,.
\ee

The same exercise can be carried out for $B_s$ mixing, where
now we allow the RS contribution to be 30\% of the SM one (that
is, $h_s^{1,4}=0.3$), without new phases~\cite{Ligeti:2010ia}.
The bounds from $Q_1$ and $Q_4$ are
\be \label{kkgluon_bs}
\begin{split}
\left( \frac{m_{\rm KK}}{2 \rm \,TeV}\right) &\gtrsim 4.7
\left(\delta f^2_{Q^{32}}\right) \approx 3\,\left(
\frac{1-2.1\,c_{Q^3}}{1-\frac23\, c_{Q^3}} \right) \,, \\
\left( \frac{m_{\rm KK}}{2 \rm \,TeV}\right) &\gtrsim 30\,
\sqrt{ \frac{\left(\delta f^2_{Q^{31}}\right)}{f_{Q^1}^2
r^g_{00}(c_{Q^1})} \left(\delta f^2_{Q^{32}}\right)
\left(\delta f^2_{D^{32}}\right) \frac{m_s}{m_b}} \,,
\end{split}
\ee
respectively. For $c_{Q^3}=0.348$ the first bound reads $m_{\rm
KK} \gtrsim 2.4$~TeV. The $Q_4$ bound is much stronger than for
$B_d$, but still weaker than the one from $Q_1$. Note that the
$Q_1$ contribution is universal, {\it i.e.}, the same for $B_d$
and $B_s\,$, and that the bound in the first line of
Eq.~\eqref{kkgluon_bs} is stronger than
Eq.~\eqref{kkgluon_bd_c1} only because we required $h_d=0.5$
and $h_s=0.3$. Eq.~\eqref{c1c4_bd} changes for $B_s$ to
\be \label{c1c4_bs}
\frac{C_4}{C_1} \Bigg|_{\textrm{2 TeV}} \approx 39\,
\frac{m_s}{m_b} \, \frac{\left(\delta f^2_{D^{32}}
\right)}{\left(\delta f^2_{Q^{32}}\right)} \left[
\left(\frac{f_{Q^3}\, r_{00}^H(\beta,c_{Q^3},c_{D^3})
}{f_{Q^2}\, r_{00}^H(\beta,c_{Q^2},c_{D^3})} \right)^2
-1\right] \,.
\ee
Note that in our example $\left(\delta f^2_{Q^{31}}\right)=
\left(\delta f^2_{Q^{32}}\right)$.

One may wonder whether $\Delta B=1$ processes, such as $b \to
s\gamma\,$, could also lead to significant bounds on the model
(see \eg~\cite{Agashe:2008uz,Gedalia:2009ws} for some recent
estimations within the anarchic scenario). However, since this
is a chirality-flipping process, it must involve right-handed
mixing angles, which are strongly suppressed in our model, as
shown in Eq.~\eqref{dr_small}. More generally, this is a
consequence of the fact that our model belongs to the class of
general MFV~\cite{Kagan:2009bn}, where right-handed currents
are suppressed by a ratio of masses, that is $m_s/m_b$ in this
case.

To summarize this example, characterized by
Eq.~\eqref{bulk_masses_ss}, the overall bound that we find is
\be
m_{\rm KK} \gtrsim 2.4 \ \mathrm{TeV}\,,
\ee
coming from the $Q_1$ contribution to $B_s$ and from EWPTs. It
should be noted that this bound can be reduced to 2.2~TeV if
compared to the SM, instead of the best fit point (with an
appropriate change in the bulk masses).

\subsection{Large 5D Bottom Yukawa}

The analysis of the previous subsection assumed a small bottom
Yukawa, as can be inferred from Eq.~\eqref{bulk_masses_ss}. Yet
by reducing $\alpha_D$, for example, the bottom Yukawa can be
made larger, until it is of order 1. Consequently, $Y_D$
resummation effects appear, and the results of the previous
subsection receive $\mathcal{O}(1)$ corrections plus a general
phase~\cite{Kagan:2009bn}.

We can try to use the large bottom Yukawa case to obtain a
larger RS contribution to $B_s$ than for $B_d$. Since $C_1$ is
universal in that sense, this requires to increase $C_4$ to be
larger than $C_1\,$, noting that $h_s^4>h_d^4$ in any case.

Considering as an example the following bulk masses
\begin{eqnarray} \label{bulk_masses_large_yb}
C_Q &=& (0.516, 0.516, 0.35)\,, \ \ \  C_U = (-0.5, -0.5, 0.479)\,, \ \ \
C_D = (0.56, 0.56, 0.497)\,,\nonumber \\
\alpha_U Y_U &=& (5.1 \times 10^{-5}, 0.025, 5.9)\,, \ \ \ \alpha_D Y_D = (0.0018, 0.034, 0.12)\,,
\end{eqnarray}
and an appropriate $\alpha_D$ to obtain a large bottom Yukawa,
we have the following results:
\begin{itemize}
\item The bound on the KK scale from EWPT is slightly
    raised to 2.6~TeV.
    \item Because of the generic phase, it is required to
        take $h_d^{1,4}$ to be 0.3 instead of
        0.5~\cite{Ligeti:2010ia}.
    \item As a result of taking $c_{D^3}=0.496$, we now
        have $h_s^4 \cong 1.33\, h_s^1\cong 0.4$, while for
        the $B_d$ system $C_4$ is still smaller than $C_1$
        (see Eqs.~\eqref{c1c4_bd} and~\eqref{c1c4_bs}, when
        evaluated at the scale 2.6~TeV originating from
        EWPT constraints).
    \item Another possible point is $c_{D^3}=0.4$ (and some
        more slight adjustments of other bulk masses). Then
        the EWPT bound is $\sim 2.7$~TeV and $h_s^4 \cong
        1.75$\,.
\end{itemize}
The implication of this result is that our model is now in
accordance with the recent Tevatron data, which favor larger
contributions to $B_s$ than for $B_d$~\cite{Ligeti:2010ia}. The
price to pay is that $\alpha_D Y_b \approx 0.12$, so that in
order to have an $\mathcal{O}(1)$ bottom Yukawa, $\alpha_D$
must be small. While this is technically natural, it still
requires a small parameter to be tuned by hand.

It is actually simple to explain why our model cannot produce
$h_s>0.3$ and $h_d\leq0.3$ if we insist on having a large
bottom Yukawa with $\alpha_D=\mathcal{O}(1)$. The latter
requirement leads to the relation $f_{Q^3} f_{D^3} \lesssim
0.01$, in order to get the correct bottom mass. However, as can
be seen from Eq.~\eqref{kkgluon_c4}, the $C_4$ contribution is
roughly proportional to $\left(f_{Q^3} f_{D^3} \right)^2$
(times another factor of $f_{Q^3}^2$ which is smaller than 1),
and as a result it is too small to yield $h_s>0.3$.

\subsubsection{The universal $h_d=h_s$ case}

While the data favor large CP violation in the $B_s$ system, a
reasonable fit of the flavor measurements is obtained in the
$SU(2)$ universal case where $h_b\equiv h_d=h_s\sim 0.3$,
consistent with the data~\cite{Ligeti:2010ia}. It is not
surprising that our framework (as well as the anarchic RS
case~\cite{Agashe:2004ay,Agashe:2004cp}) can account for this
case in a straightforward manner, while having $\alpha_D$ and
the 5D bottom Yukawa of order unity. This is obtained by taking
$c_{D^3}$ to be $\sim$0.6 and $c_{D^i}$ around 0.6-0.65, while
the other bulk masses are as in Eq.~\eqref{bulk_masses_ss}. In
this case, one can sharply predict order 1 CPV phases with
exact universality, $\sigma_b\equiv\sigma_d=
\sigma_s$~\cite{Kagan:2009bn}. The resulting EWPT bound is
$\sim2.4$~TeV.

\subsection{Higgs Mediated FCNCs}

Another possible source of flavor violation arises from the
Higgs~\cite{Agashe:2009di,Azatov:2009na,Duling:2009pj}, which
obtains off-diagonal couplings in the mass basis as a result of
mixing between zero-mode and KK-fermions. For an IR brane
Higgs, the leading spurion which induces this process
is~\cite{Azatov:2009na}
\begin{equation} \label{higgs_fv}
\sim F_Q Y_D Y_D^\dagger Y_D F_D^\dagger \,,
\end{equation}
omitting all universal factors\footnote{An additional
contribution comes from a one-loop process involving a charged
Higgs and up-type quarks. However, as a result of the loop
suppression, it is subleading.}. Yet, the resulting flavor
violation is suppressed relative to the KK-gluon contribution.
To see this, let us neglect the masses (and Yukawa couplings)
of quarks of the first two generations. Then in its diagonal
basis, $Y_D$ is proportional to diag$(0,0,Y_b)$, and
consequently we have $Y_D^3 \propto Y_b^2 Y_D$. In other words,
the leading mass term in Eq.~\eqref{mass0} and the spurion in
Eq.~\eqref{higgs_fv} are aligned together, so no flavor
violation is generated. Restoring the strange mass, we expect
to have a $(m_s/m_b)^2$ suppression, after squaring these
spurions to obtain the relevant Wilson coefficients. Since a
factor of this kind does not appear for the KK-gluon
contribution to flavor violation via $Q_1$, the Higgs effect
can be neglected.

This argument is easily generalized to the bulk Higgs case. The
$Y_D^3$ part of Eq.~\eqref{higgs_fv} should be written now as
\begin{equation}
Y_D r_{01}^H Y_D^\dagger r_{10}^H Y_D \,,
\end{equation}
where $r^H_{01,10}$ is an overlap correction for the coupling
of the Higgs to a zero-mode quark and a KK-quark. Even though
these corrections are not universal, the wrapping $Y_D$'s act
as a projection operator for the 3-3 matrix element, when
neglecting the first two generations' masses. Therefore, we
still have $Y_D^3 \propto Y_b^2 Y_D\,$, and the conclusion from
before applies to this case as well. Moreover, we did not have
to assume anything about $F_Q$ and $F_D$, hence the Higgs
contribution is negligible in both the small and the large
bottom Yukawa cases.

\begin{table}[htb]
\begin{center}
\begin{tabular}{|l|c|c|c|c|}
\hline & \multicolumn{2}{|c|}{One d.o.f.} & \multicolumn{2}{|c|}{Six/three d.o.f.} \\
\hline Model & \ Best fit \ & \ \ \ \ SM \ \ \ \ & \ Best fit \ & \ \ \ \ SM \ \ \ \ \\
\hline Flavor triviality & 2.1 & 1.8 & 1.7 & 1.6 \\
\hline Semi-anarchy & 7 & 4.6 & 4.6 & 3.9 \\ \hline
\end{tabular}
\end{center}
\caption{Bounds (in TeV) from EWPTs for the various statistical
scenarios considered, comparing the flavor triviality model to
the semi-anarchic case. Best fit refers to the bound relative
to the best fit point, where $\chi^2$ is lower than in the SM,
and SM refers to the case where the SM is assumed to be the
minimum $\chi^2\,$. In the last two columns the flavor
triviality analysis is for 6 d.o.f., while the semi anarchic
one is for 3 d.o.f.} \label{sum_bounds}
\end{table}

\section{Conclusions} \label{sec:conc}

We analyzed a warped 5D model where the SM Yukawa hierarchy is
set by UV physics, which realizes a bulk version of the
Rattazzi-Zaffaroni model~\cite{Rattazzi:2000hs}. Such a
scenario displays the weakest bound on the scale of RS type of
new physics explored to date, both from the point of view of
electroweak precision measurements, as well as from flavor
constraints. The EW precision tests allow for a ``sweet spot''
with a KK scale as low as 2.1~TeV, which is more than a factor
of 2 lower than in the anarchic RS setup (with fermions in the
minimal $SU(2)_L \times SU(2)_R$ representations, as discussed
in the main text). Such a low scale for the RS KK physics
should lead to significantly better prospects for discovery at
the LHC, given the fact that its reach for a KK-gluon is around
4~TeV~\cite{Agashe:2006hk,Lillie:2007yh}. A summary of the
bounds that we find from EWPTs for different statistical
scenarios is presented in Table~\ref{sum_bounds}.

This model, by construction, belongs to the minimal flavor
violation (MFV) framework~\cite{Chivukula:1987py,
Hall:1990ac,Gabrielli:1994ff,
Ali:1999we,Buras:2000dm,D'Ambrosio:2002ex,Buras:2003jf,
Buras:2005xt,Hurth:2008jc,Isidori:2009px}, and naively one
would expect a rather dull flavor phenomenology. Indeed the
model is strongly protected from CP violation in the first two
generations. Imposing the flavor constraints, we also find
consistency with a KK scale of about $2.4$~TeV. Thus the RS
$\epsilon_K$ problem is avoided, practically, without
interfering with the model's visibility. This is a natural
consequence of MFV. However, the fact that in this framework
the third generation couplings are sizable and flavor-violating
couplings effectively exponentiate leads to various interesting
deviations from the commonly studied MFV models. Thus, this
class of models belongs to the \textit{general} MFV
framework~\cite{Kagan:2009bn}. Performing a deformation around
the best fit point in parameter space allows for a rather rich
third generation flavor phenomenology, such as providing the
new CPV source required by the latest same-sign di-muon signal
from D$\slashed{0}$. The present ideas are expected to help
reduce the constraints (hence the fine-tuning) of
extra-dimensional scenarios of EWSB, such as models where the
Higgs is a pseudo-Goldstone boson. The detailed study of such
an exciting possibility is left to future work.

\vspace{3mm}
\mysection{Acknowledgements}

We are grateful to Kaustubh Agashe and Jos\'e Santiago for
valuable discussion, comments on the manuscript and helping us
to find an error in our global fit. GP is the Shlomo and Michla
Tomarin career development chair; GP is supported by the Israel
Science Foundation (grant \#1087/09), EU-FP7 Marie Curie, IRG
fellowship and the Peter \& Patricia Gruber Award. E.P. is
supported by DOE grant DE-FG02-92ER40699.

\appendix

\section{Matching RS to the EW precision operators}
\label{app:ew}

New physics effects at the weak scale are captured by a set of
effective operators added to the renormalizable part of the SM
Lagrangian: $\CL=\CL_{\rm SM}+\sum_ia_i\CO_i\,$, where $\CO_i$
are gauge and flavor invariant operators. In the absence of
flavor and CP violation, 20 operators~\footnote{An additional
operator, $\CO_W=\epsilon_{abc} W_{\mu\nu}^a W^{\nu\rho b}
W_\rho^{\mu c}$, can be considered as well. However, it is
weakly constrained by EWPT, since it affects only the triple
and quadruple gauge self-couplings, which are poorly measured.
Thus we set this operator to zero in our fit.} (of mass
dimension~6) contribute most significantly to the electroweak
precision observables~\cite{Han:2004az}. There are 2 operators
affecting the gauge sector,
\beq
\CO_{WB}=h^\dagger \sigma^a h W_{\mu\nu}^aB^{\mu\nu}\,,\quad
\CO_{h}=|h^\dagger D_\mu h|^2\,,
\eeq
which generate, respectively, the $S$ and $T$ parameters, 7
operators shifting the fermion-gauge boson couplings
\be\label{eqn:O6hf}
\begin{split}
\CO_{hf}^{s}&=ih^\dagger D_\mu h \bar{f}\gamma^\mu f + \hc \,,\quad
\CO_{hF}^s=ih^\dagger D_\mu h \bar{F}\gamma^\mu F + \hc \,,\\
\CO_{hF}^t&=ih^\dagger \sigma^aD_\mu h \bar{F}\gamma^\mu
\sigma^a F+\hc\,,
\end{split}
\ee
where $f=u,d,e$ and $F=q,l$, and 11 four-fermion operators
contributing to the leptonic sector
\be \label{eqn:O6lq2}
\begin{split}
\CO_{ll}^s&=\half (\bar{l}\gamma^\mu l)^2\,,\ \CO_{ll}^t=\half
(\bar{l}\gamma^\mu\sigma^a l)^2,\ \CO_{le}^s=(\bar{l}\gamma^\mu
l)(\bar{e}\gamma_\mu e)\,,\ \CO_{ee}^s=\half(\bar{e}\gamma^\mu e)^2\,,\\
\CO_{lq}^s&=(\bar{l}\gamma^{\mu} l)(\bar{q}\gamma_{\mu} q)\,,\
\CO_{lq}^t=(\bar{l}\gamma^{\mu}\sigma^a l)(\bar{q}\gamma_{\mu}
\sigma^a q)\,,\ \CO_{qe}^s=(\bar{q}\gamma^{\mu} q)(\bar{e}\gamma_{\mu} e)\,,\\
\CO_{lu}^s&=(\bar{l}\gamma^\mu l)(\bar{u}\gamma_\mu u)\,,\
\CO_{ld}^s= (\bar{l}\gamma^\mu l)(\bar{d}\gamma_\mu d)\,,\
\CO_{eu}^s=(\bar{e}\gamma^\mu e) (\bar{u}\gamma_\mu u)\,,\
\CO_{ed}^s=(\bar{e}\gamma^\mu e)(\bar{d}\gamma_\mu d)\,.
\end{split}
\ee
Whenever relevant, a U(3) trace over flavor is assumed in all
of the above. Given the peculiar behavior of the third
generation quarks, new physics is expected to break the
U(3)$^3$ flavor symmetries in the quark sector down to
$[$U(2)$\times$U(1)$]^3$. In our setup $b_R$ behaves as the
lighter generations $d_R$ and $s_R\,$, and EWPTs are not
sensitive to top observables. Thus, it is well justified to
work in a limit where only ${\rm U}(3)_Q$ is broken down to
${\rm U}(2)_{q}\times {\rm U}(1)_{Q}\,$, with $q$ and $Q$
denoting the first two and third generation quark doublets,
respectively. In this case there are 5 additional operators
\be
\begin{split}
\CO_{hQ}^s &= ih^\dagger D_\mu h \bar{Q}\gamma^\mu Q + \hc\, ,\
\CO_{hQ}^t=ih^\dagger \sigma^a D_\mu h \bar{Q}\gamma^\mu
\sigma^a Q+\hc\,,\\ \CO_{lQ}^t &= (\bar{l}\gamma^\mu \sigma^a
l) (\bar{Q}\gamma_\mu\sigma^a Q)\,,\
\CO_{lQ}^s=(\bar{l}\gamma^\mu l) (\bar{Q}\gamma_\mu Q)\,,\
\CO_{Qe}^s=(\bar{Q}\gamma^\mu Q)(\bar{e}\gamma_\mu e)\,,
\end{split}
\ee
and a U(2) trace over the first two generations is now
understood in the $\bar{q}\gamma^\mu q$ current of
$\CO_{hq}^{s,t}$, $\CO_{lq}^{s,t}$ and $\CO_{qe}^s$. Therefore,
25 operators are relevant for EWPT. We use the SM global fit
of~\cite{PDG2010}, following the approach developed
in~\cite{Han:2004az, Han:2005pr}, with updated $m_{\rm top} =
(173.3 \pm 1.1)~{\rm
GeV}$~\cite{TevatronElectroweakWorkingGroup:2010yx} and $m_{W}
= (80.420 \pm 0.031)~{\rm GeV}$ ~\cite{:2009nu} measurements
from the Tevatron.

\paragraph{Tree-level effects}
We start with matching the coefficients of the 25 operators to
RS at tree-level. The leading contributions arise from exchange
of gauge KK-modes, as depicted in the diagrams of
Fig.~\ref{fig:diagtree}. An explicit evaluation of the latter
yields (see~\cite{Davoudiasl:2009cd} for a pedagogical
review~\footnote{For reference, the relation between the
notation used here and that of \cite{Davoudiasl:2009cd} is as
follows: $\alpha^{N} = L \, G_{++}$, $\alpha^{D} = L \,
G_{-+}$, $\beta^{N}_{\psi} = L \, I_{++}(c_{\psi})$,
$\beta^{D}_{\psi} = L \, I_{-+}(c_{\psi})$,
$\gamma^{N}_{\psi\psi'} = L \, J_{++}(c_{\psi},c_{\psi'})$ and
$\gamma^{D}_{\psi\psi'} = L \, J_{-+}(c_{\psi},c_{\psi'})$,
where $L = R \log[R'/R]$ is the proper size of the fifth
dimension.})
\begin{eqnarray}
\label{as}
a_h&=&\frac{g_5'^2}{2}(G_{++}-G_{-+})\,,\nonumber\\
a_{hF}^t&=&\frac{g_5^2}{4}I_{++}(c_F)\,,\nonumber\\
a_{FF'}^t&=&\frac{g_5^2}{4}J_{++}(c_F,c_{F'})\,,\nonumber\\
a_{hF}^s&=&\frac{g_5'^2}{2}Y_F\left[I_{++}(c_F)-I_{-+}(c_F)\right]+\frac{g_{5R}^2}{2}T^F_{3R}I_{-+}(c_F)\,,\\
a_{hf}^s&=&\frac{g_5'^2}{2} Y_f \left[I_{++}(c_{f})-I_{-+}(c_{f})\right]+\frac{g_{5R}^2}{2}T^f_{3R}I_{-+}(c_{f})\,,\nonumber\\
a_{FF'}^s&=&g_5'^2Y_FY_{F'}J_{++}(c_F,c_{F'})+\frac{g_{5R}^2}{\cos^2\theta}\left(T_{3R}^F-\sin^2\theta Y_F\right)
\left(T_{3R}^{F'}-\sin^2\theta Y_{F'}\right)J_{-+}(c_F,c_{F'})\,,\nonumber\\
a_{ff'}^s&=&g_5'^2Y_fY_{f'}J_{++}(c_{f},c_{f'})+\frac{g_{5R}^2}{\cos^2\theta}\left(T_{3R}^f-\sin^2\theta Y_f\right)\left(T_{3R}^{f'}
-\sin^2\theta Y_{f'}\right)J_{-+}(c_{f},c_{f'})\,,\nonumber
\end{eqnarray}
where $F,F'=Q,q,l$ and $f,f'=u,d,e$ with
$\sin^2\theta=g_5'^2/g_{5R}^2$. The $G,I,J$ wave-function
overlap integrals are given by
\begin{eqnarray}
G_{\pm+}&=&v^{-4}\int_R^{R'}dz dz'\left(\frac{R}{z}\right)^3\left(\frac{R}{z'}\right)^3
v_5(z,\beta)^2\mathcal{G}_{\pm+}(z,z') v_5(z',\beta)^2\,,\nonumber\\
I_{\pm+}(c)&=&v^{-2}\int_R^{R'}dz dz'\left(\frac{R}{z}\right)^4\left(\frac{R}{z'}\right)^3
\chi(z,c)^2\mathcal{G}_{\pm+}(z,z') v_5(z',\beta)^2\,,\\
J_{\pm+}(c,c')&=&\int_R^{R'}dz dz'\left(\frac{R}{z}\right)^4\left(\frac{R}{z'}\right)^4
\chi(z,c)^2\mathcal{G}_{\pm+}(z,z') \chi(z',c')^2\,,\nonumber
\end{eqnarray}
where $\mathcal{G}_{\pm+}$ is the mixed position-momentum 5D
propagator for $(\pm,+)$ gauge bosons in AdS space evaluated at
zero (4D) momentum~\cite{Randall:2001gb,Carena:2003fx} and
$\chi(z,c)$ is the fermion zero-mode wave-function, while
$v_5(z,\beta)$ is the bulk Higgs VEV. $g_5$, $g_5'$ and
$g_{5R}$ are the 5D gauge couplings of SU(2)$_L$, U(1)$_Y$ and
SU(2)$_R\,$, respectively. While $g_5$ and $g_5'$ have to be
matched to the 4D gauge couplings (see below), $g_{5R}$ is a
free parameter of the model, which we take to be
$g_{5R}=g_5\,$, as required by the extended custodial symmetry
that protects the $Zb_L\bar{b}_L$ coupling. Note that
$a_{WB}\sim \CO(v^4/m_{\rm KK}^4)$ at tree-level, and we recall
that the tree-level $S$ parameter often quoted in RS is coming
from a ``universal'' shift to the fermion couplings. This
contribution is included in the global fit through the shifts
in the fermion-gauge boson couplings, which is just a
consequence of the fact that some operators in the effective
Lagrangian are
redundant~\cite{Grojean:2006nn,Cacciapaglia:2006pk}.

\paragraph{Matching of 5D gauge couplings}

The 5D gauge couplings used above have to be matched to their
4D values in the effective action. Including one-loop
renormalization, the matching conditions
are~\cite{Pomarol:2000hp,Randall:2001gb, Randall:2001gc,
Goldberger:2002cz,Agashe:2002bx, Contino:2002kc,
Goldberger:2002hb, Choi:2002ps, Agashe:2002pr,
Goldberger:2003mi}
\beq
\frac{1}{g^{2}}=\log(R'/R)\left(\frac{1}{g_5^{2}k}
+\frac{b_{g}}{8\pi^2}\right) +\frac{1}{g_{\rm UV}^{2}}
+\frac{1}{g_{\rm IR}^{2}}\,,
\eeq
where the last two terms are contributions from (possible)
``bare'' brane-localized kinetic terms, which we set to zero for
simplicity. The one-loop $\beta$-function coefficients $b$
receive contributions from the bulk only through elementary
fields. Hence after removing the Higgs contribution from the
running, we find $b_g=-10/3$ and $b_{g'}=20/3$. Therefore,
matching the 5D gauge couplings at the TeV scale yields
$g_5\sqrt{k}\simeq 27.3g/\sqrt{\log(R'/R)}\simeq 3.02$ and
$g_5'\sqrt{k}=43.9g'/\sqrt{\log(R'/R)}\simeq 2.66$ for
$k=R^{-1}=2.4 \times 10^{18}\,$GeV and $m_{\rm KK}\sim 2\,$TeV.

\paragraph{One-loop effects}
The large top Yukawa induces a non-negligible contribution to
the S and T parameters, as well as to the $b_L$ coupling to Z.
A straightforward calculation of the one-loop diagram of
Fig.~\ref{fig:diagoneloop} gives the following contributions to
the oblique parameters~\cite{Lavoura:1992np,Carena:2006bn}
\begin{eqnarray}
S&=&\frac{N_c}{2\pi}\sum_{\alpha,\beta}\sum_{X=U,D}\Big[
\left(X^{L\dagger}_{\alpha\beta}Y^L_{X\beta\alpha}+X^{R\dagger}_{\alpha\beta}
Y^R_{X\beta\alpha}\right)\chi_+(m_X^\alpha,m_X^\beta)\nonumber\\
&&\hspace{3cm}+\left(X^{L\dagger}_{\alpha\beta}Y^R_{X\beta\alpha}
+X^{R\dagger}_{\alpha\beta}Y^L_{X\beta\alpha}\right)\chi_-(m_X^\alpha,m_X^\beta)\Big]\,,\nonumber\\
T&=&\frac{N_c}{16\pi s_W^2c_W^2m_Z^2}\Big[\sum_{\alpha,i}V_{\alpha i}^2\theta_+
(m_U^\alpha,m_D^i)\nonumber +A^V_{\alpha i}\theta_-(m_U^\alpha,m_D^i)\nonumber\\
&&\hspace{3cm}-\sum_{\beta <\alpha}U_{\alpha \beta}^2\theta_+(m_U^\alpha,m_U^\beta)
+A^U_{\alpha\beta}\theta_-(m_U^\alpha,m_U^\beta)\Big]\,,\label{eqn:ST1loop}
\end{eqnarray}
where we defined $K^2\equiv |K^L|^2+|K^R|^2$, $A^K=2\,{\rm
Re}\left[K^LK^{R*}\right]$ with $K=U,V$. The unitary matrices
$U^{L,R}$,$D^{L,R}$ ($Y_{U,D}^{L,R}$) denote the couplings of
the $Q=2/3$ and $Q=(-1/3,5/3)$ mass eigenstates to the
$W_{3L}^\mu$ ($B^\mu$) zero-mode, while the $V^{L,R}$ matrices
stand for the coupling of the mass eigenstates to the $W^\pm$
zero-mode. The definitions of the loop functions $\theta_{\pm}$
and $\chi_{\pm}$ are~\cite{Lavoura:1992np}
\beq
\theta_+(m_1,m_2)&=&m^2_1+m_2^2-\frac{2m_1^2m_2^2}{m_1^2-m_2^2}\log\frac{m_1^2}{m_2^2}\,,\\
\theta_-(m_1,m_2)&=&2m_1m_2\left(\frac{m_1^2+m_2^2}{m_1^2-m_2^2}\log\frac{m_1^2}{m_2^2}-2\right)\,,\\
\chi_+(m_1,m_2)&=&\frac{5(m_1^4+m_2^4)-22m_1^2m_2^2}{9(m_1^2-m_2^2)^2}-\frac{2}{3}\log\frac{m_1m_2}{\mu^2}\nonumber\\
&&+\frac{3m_1^2m_2^2(m_1^2+m_2^2)-(m_1^6+m_2^6)}{3(m_1^2-m^2_2)^3}\log\frac{m_1^2}{m_2^2}\,,\\
\chi_-(m_1,m_2)&=&\frac{m_1m_2}{(m_1^2-m_2^2)^3}\left(m_1^4-m_2^4-2m_1^2m_2^2\log\frac{m_1^2}{m_2^2}\right)\,,
\eeq
where the renormalization scale dependence in $\chi_+$ cancels
out in $S$ thanks to tr$[U^\dagger Y_U+D^\dagger Y_D]=0$. Note
that Eq.~\eqref{eqn:ST1loop} includes SM contributions from top
and bottom
\beq
S_{\,\rm SM}\simeq \frac{N_c}{18\pi}
\left[3-\log\left(\frac{m_t^2}{m_b^2}\right)\right],\ T_{\rm
SM}\simeq \frac{N_c}{16\pi s_W^2c_W^2}
\left(\frac{m_t^2}{m_Z^2}\right)\,,
\eeq
which need to be subtracted in order to isolate the new physics
contributions.

These loop effects are controlled by EWSB, dominantly from the
top sector as parametrized by the 5D top Yukawa. The
contributions associated with EWSB mixing among heavy KK-modes
are controlled by the top Yukawa coupling evaluated at a scale
of the order of the KK masses. These contributions are
subdominant, however, and the result is dominated by EW mixing
between the KK-modes and the top zero-mode. The relevant
diagrams display an IR divergence that is cutoff by the top
mass, indicating that the result is dominated by scales of
order $\mu \sim m_{\rm top}$ (see~\cite{Carena:2006bn}). To be
conservative, we use in these loop contributions a top running
mass $m_{t}(\mu = m_{\rm top}) \approx 160~{\rm
GeV}$~\cite{Xing:2007fb}. Similarly, we use $m_{b}(\mu = m_{\rm
top}) \approx 2.7~{\rm GeV}$.

In gauge-Higgs unified models, where the Higgs is realized as a
pseudo-Goldstone boson, the Higgs only partially regulates the
divergent contribution to the $S$ and $T$ parameters arising
from loops of (longitudinal) SM gauge fields\footnote{We thank
Kaustubh Agashe for bringing this point to our
attention.}~\cite{Barbieri:2007bh}. This results in a
logarithmic correction to the $S$ and $T$ parameters which is
cut by the KK scale~\cite{Contino:2010rs},
\beq\label{eqn:SThiggs}
\Delta S=\frac{1}{12\pi} \left(1-a^2\right)\log
\frac{\Lambda_{\rm eff}^2}{m_h^2}\,,\quad \Delta T=-\frac{3}{16\pi
c_W^2}\left(1-a^2\right)\log\frac{\Lambda_{\rm eff}^2}{m_h^2}\,,
\eeq
where $\Lambda_{\rm eff}\simeq m_{\rm KK}$ and $a$ measures the
amount of Higgs compositeness, with $a=1$ corresponding to the
fully elementary SM Higgs. Deviation from $a=1$ also leads to
an incomplete unitarization of the W/Z scattering amplitude,
and perturbative unitarity is lost at $\Lambda_{\rm eff}\simeq
1.2$ TeV$/\sqrt{|1-a^2|}$. Requiring unitarity not to be
violated below the KK scale, we find the contributions
in~\eqref{eqn:SThiggs} to raise the bound for one (six) d.o.f.
by 300 (200)~GeV, assuming the SM as the best fit.

The $Z\bar{b}_Lb_L$ vertex also receives large radiative
corrections dominated by the diagrams of
Fig.~\ref{fig:diagZbbloop}. This yields
\beq
\delta g^{b_L}&=&\frac{\alpha}{2\pi}\Bigg[\sum_{\alpha}V_{\alpha
b}^L V_{\alpha b}^L\left[F_{\rm SM}(r_\alpha)+\tilde{F}
\left(\frac{U_{\alpha\alpha}^L}{2}
-\half,\frac{U_{\alpha\alpha}^R}{2},r_\alpha\right)\right]\nonumber\\
&&\hspace{2cm}+\sum_{\alpha<\beta}V_{\alpha b}^LV_{\beta
b}^L\mathcal{F}
\left(\frac{U_{\alpha\beta}^L}{2},\frac{U_{\alpha\beta}^R}{2},
r_\alpha,r_\beta\right)\Bigg]\,,
\eeq
where $r_\alpha=(m_U^\alpha/m_W)^2$ and the loop functions
are~\cite{Bamert:1996px,Carena:2007ua}
\beq
F_{\rm SM}(r)&=&\frac{r}{8s_W^2}\frac{(r-1)(r-6)+(3r+2)\log r}{(r-1)^2}\,,\\
\tilde{F}(g_L,g_R,r)&=&\frac{r}{8s_W^2}\Big[g_L\left(2-\frac{4}{r-1}\log r\right)\nonumber\\
&&-g_R\left(\frac{2r-5}{r-1}+\frac{r^2-2r+4}{(r-1)^2}\log r\right)\Big]\,,\\
\mathcal{F}(g_L,g_R,r,r')&=&\frac{1}{4s_W^2(r'-r)}\Big[2g_L\left(
\frac{r-1}{r'-1}r'^2\log r'-\frac{r'-1}{r-1}r^2\log r\right)\nonumber\\
&&-g_R\sqrt{rr'}\left(r'-r+\frac{r'-4}{r'-1}r'\log
r'-\frac{r-4}{r-1}r\log r\right)\Big]\,.
\eeq
Here again the SM contribution,
\beq
\delta g^{b_L}_{\rm SM}=\frac{\alpha}{2\pi}F_{\rm SM}(r_t)\,,
\eeq
should be subtracted to isolate the contribution from new
physics. Note that this result is derived for an off-shell
($q^2=0$) Z in the 't Hooft/Feynman ($\xi=1$) gauge. Although
the result should only be gauge-invariant when the Z is
on-shell ($q^2=m_Z^2$), we expect the missing terms to suffer
an additional $m_Z^2/m_{\rm KK}^2$ suppression, so the $q^2=0$
result quoted constitutes a valid approximation for the new physics
contribution. Notice also that all the radiative corrections
above decouple like $v^2/m_{\rm KK}^2$, as expected, since they
arise from vector-like (KK-)fermions which mix with the chiral
zero-mode through Yukawa couplings.

The above one-loop corrections are accounted for in the global
fit by adding the following shifts to the coefficients of the
$\CO_{WB}$, $\CO_{h}$ and $\CO_{hQ}^s$ operators:
\beq
a_{WB}&\to& a_{WB}+\frac{gg'}{16\pi v^2}(S-S_{\,{\rm SM}})\,,
\nonumber\\
a_{h} &\to& a_h -\frac{g^2 s_W^2}{2\pi v^2}(T-T_{\rm SM})\,,\\
a_{hQ}^s&\to& a_{hQ}^s-\frac{2}{v^2}(\delta g^{b_L}-\delta
g^{b_L}_{\rm SM})\,.
\nonumber
\eeq
As for $S$ and $T\,$, we use a renormalization scale of order
$\mu \sim m_{\rm top}$ to evaluate $\delta g^{b_L}$, which errs
on the conservative side.

\section{Contributions to B meson mixing} \label{app:flavor}

In Sec.~\ref{sec:flavor} we estimated the bounds coming from
the contributions to $B_{d,s}$ mixing in our model. Here we
calculate these contributions in detail. We begin with the
simpler case of small bottom Yukawa coupling, based on the bulk
masses of Eq.~\eqref{bulk_masses_ss}, and then we generalize to
the large bottom Yukawa case, and show that in fact there are
only $\CO$(1) corrections.

\subsection{Small 5D Bottom Yukawa}

We start with the mass relation of Eq.~\eqref{mass0}, where for
simplicity we omit the overlap correction $r^H_{00}$ (which can
be restored at the end) and absorb the $\alpha_{U,D}$
coefficients into the 5D Yukawas. As explained in
Sec.~\ref{sec:small_yb}, we have to a good approximation
$\left[m_U,Y_U \right]=0$. Thus, it is convenient to work in a
basis in which $Y_U$ is diagonal. The 5D Yukawa matrices can
then be written as
\be
Y_U=\lambda_U\,, \quad Y_D=V^Q \lambda_D\,,
\ee
where $\lambda_U=\mathrm{diag}(Y_u,Y_c,Y_t)\,$,
$\lambda_D=\mathrm{diag}(Y_d,Y_s,Y_b)$ and $V^Q$ is the
misalignment between $Y_U$ and $Y_D$, or in other words, the 5D
equivalent of the CKM matrix.

In order to find the relation between $V^Q$ and $\vckm$, we
note that the latter diagonalizes the mass matrix $m_D$ from
the left, {\it i.e.} it diagonalizes
\be \label{md_left1}
m_D m_D^\dagger \propto F_Q Y_D F_D F_D^\dagger Y_D^\dagger
F_Q^\dagger \,.
\ee
The almost universal $F_D$'s can be thrown away, since we now
only care about diagonalization, and for the $F_Q$'s we can
pull out a factor of $f_{Q^1}$ ($=f_{Q^2}$), obtaining
\be
F_Q \propto \rqmat \,,
\ee
where we defined\footnote{To be precise, the right-handed bulk
mass that is used in the overlap corrections in Eq.~\eqref{rq}
depends on the process in which we use $r_Q\,$. Since the
largest contributions usually come from the third generation,
we defined $r_Q$ with $c_{D^3}\,$.}
\be \label{rq}
r_Q \equiv \frac{f_{Q^3}\, r^H_{00}(\beta, c_{Q^3},
c_{D^3})}{f_{Q^1}\, r^H_{00}(\beta, c_{Q^1}, c_{D^3})} \,.
\ee
Eq.~\eqref{md_left1} then becomes
\be \label{md_left2}
m_D m_D^\dagger \propto \rqmat\, V^Q \, \lambda_D^2\,
V^{Q\dagger}\, \rqmat \,.
\ee
From this expression, it is simple to find the following
relations for the mixing angles\footnote{We omit any complex
conjugate signs here and below.}
\be \label{vq_to_vckm}
V^Q_{12} \sim \vckm_{us} \,, \quad V^Q_{13} \sim  r_Q
\vckm_{td} \,, \quad V^Q_{23} \sim r_Q \vckm_{ts} \,.
\ee
The 5D CKM mixing angles for the third generation are thus
larger than the corresponding CKM elements. This is not a
surprise, since the hierarchy in the 5D Yukawas is milder than
for the masses because of the $f_Q$'s. After diagonalization,
we find the mass relations
\be \label{md_relations}
m_{d,s,b} \cong \frac{v}{\sqrt2} f_{Q^{1,1,3}} Y_{d,s,b}
f_{D^1} \,,
\ee
where we used the facts that $f_{Q^1}=f_{Q^2}$ in our current
realization of the model and that all the $f_D$'s are almost
identical.

The matrix $D_R$ which diagonalizes $m_D$ from the right is
computed from the following expression:
\be \label{md_right}
m_D^\dagger m_D \propto F_D^\dagger Y_D^\dagger F_Q^\dagger F_Q
Y_D F_D \propto \lambda_D\, V^{Q\dagger} \rqsmat\, V^Q \,
\lambda_D \,.
\ee
The resulting mixing angles of $D_R$ are
\be \label{dr_small}
\begin{split}
\left(D_R\right)_{12} &\sim (r_Q^2-1)\, \frac{Y_d}{Y_s} \,
V^Q_{13}\,V^Q_{23} \sim (r_Q^2-1) r_Q^2\,
\frac{m_d}{m_s} \, \vckm_{td}\, \vckm_{ts} \,, \\
\left(D_R\right)_{13} &\sim \frac{r_Q^2-1}{r_Q^2} \,
\frac{Y_d}{Y_b} \,V^Q_{13} \sim (r_Q^2-1)\, \frac{m_d}{m_b} \,
\vckm_{td} \,, \\ \left(D_R\right)_{23} &\sim
\frac{r_Q^2-1}{r_Q^2} \, \frac{Y_s}{Y_b} \, V^Q_{23} \sim
(r_Q^2-1)\, \frac{m_s}{m_b} \, \vckm_{ts} \,,
\end{split}
\ee
where we used Eqs.~\eqref{vq_to_vckm} and~\eqref{md_relations}.

The operators considered in Sec.~\ref{sec:flavor} are generated
by KK-gluon exchange. The coupling of two zero-mode quarks to
a KK-gluon is proportional to $F_Q F_Q^\dagger$ or $F_D^\dagger
F_D$ for left- or right-handed couplings, respectively. Applying
the appropriate rotation to the down mass basis, the 1-3 entries of
the couplings, relevant for $B_d$ mixing, are
\be \label{fq_fd_13}
\begin{split}
\left(F_Q F_Q^\dagger\right)_{13} \Big|_{\textrm{mass basis}}
~\sim~& \vckm_{tb} \left(\vckm_{td}\right)^* \left[ f_{Q^3}^2
r_{00}^g(c_{Q^3}) -f_{Q^1}^2 r_{00}^g(c_{Q^1}) \right]\,, \\
\left(F_D^\dagger F_D\right)_{13} \Big|_{\textrm{mass basis}}
~\sim~& \vckm_{tb} \left(\vckm_{td}\right)^* \frac{m_d}{m_b}
\left[ \left(\frac{f_{Q^3}\, r_{00}^H(\beta,c_{Q^3},c_{D^3})
}{f_{Q^1}\, r_{00}^H(\beta,c_{Q^1},c_{D^3})} \right)^2
-1\right] \\ &\times \left[ f_{D^3}^2 r_{00}^g(c_{D^3})
-f_{D^1}^2 r_{00}^g(c_{D^1}) \right] \,.
\end{split}
\ee
The result for $B_s$ is obtained by the replacements $1 \to 2$
and $d \to s\,$.

It should be noted that $\left(D_R\right)_{12}$ from
Eq.~\eqref{dr_small} is actually not useful, since when a $D_R$
rotation is applied to $F_D^\dagger F_D\,$, the 1-2 entry is
multiplied by $\sim (f_{D^2}^2-f_{D^1}^2)$, which is zero in
our model. Hence, FCNC processes among the first two
generations follow through $\left(D_R\right)_{13} \cdot
\left(D_R\right)_{23}\,$. This explains why the right-handed
current for $\epsilon_K$ is suppressed by $r_Q^4 m_d m_s/m_b^2$
relative to the left-handed current, as mentioned in
Sec.~\ref{sec:flavor}.

\subsection{Large 5D Bottom Yukawa}

The case where the 5D bottom Yukawa is large is more
complicated, but it turns out that it only leads to $\CO$(1)
corrections, as we now show. First, $Y_U$ and $m_U$ do not
commute anymore, so there is no ``natural'' basis to adopt. It
is therefore useful to define two new matrices, $V^{QD,QU}$,
which parametrize the misalignment between $Y_{D,U}$ and $C_Q$
(and equivalently $F_Q$). Moreover, we need now to compute both
$D_L$ and $U_L$ (diagonalizing $m_D$ and $m_U$ from the left,
respectively), in order to relate all the above matrices to the
CKM matrix. In the following we consider for simplicity only
the first relevant terms in the MFV expansion of the 5D
spurions.

The first step is to relate $V^{QD}$ and $V^{QU}$ to $V^Q$. In
the basis in which $Y_D$ is diagonal, $C_Q$ from
Eq.~\eqref{bulk_masses_short} can be written as
\be
C_Q=a_Q\cdot \mathbf{1}_3+b_D^Q \,\lambda_D^2 +b_U^Q
\,V^{Q\dagger} \lambda_U^2 V^Q+\dots\,,
\ee
and it is diagonalized by $V^{QD}$. We then obtain the
following relations:
\be \label{VQD}
V^{QD}_{12} \sim \frac{V^Q_{13}}{V^Q_{23}}\sim V^Q_{12}\,,
\quad V^{QD}_{13} \sim V^Q_{13} \left( \frac{b_U^Q Y_t^2}{b_D^Q
Y_b^2 +b_U^Q Y_t^2} \right)\,, \quad V^{QD}_{23} \sim V^Q_{23}
\left( \frac{b_U^Q Y_t^2}{b_D^Q Y_b^2 +b_U^Q Y_t^2} \right)\,,
\ee
where we assumed a specific relation between the $V^Q$ mixing
angles for $V^{QD}_{12}\,$, which is consistent with the
results below (since a similar relation holds for the CKM
matrix). Note that the expression in parentheses in the last
two mixing angles is of order 1 as long as $Y_b$ is smaller
than or comparable to $Y_t$ and $b^Q_{U,D}$ are $\CO$(1).
Similarly, in the basis in which $Y_U$ is diagonal, $C_Q$ can
be written as
\be
C_Q=a_Q\cdot \mathbf{1}_3+b_U^Q \,\lambda_U^2 +b_D^Q \,V^Q
\lambda_D^2 V^{Q\dagger}+\dots\,.
\ee
We then have
\be \label{VQU}
V^{QU}_{12} \sim V^Q_{12}\,, \quad V^{QU}_{13} \sim V^Q_{13}
\left( \frac{b_D^Q Y_b^2}{b_D^Q Y_b^2 +b_U^Q Y_t^2} \right)\,,
\quad V^{QU}_{23} \sim V^Q_{23} \left( \frac{b_D^Q Y_b^2}{b_D^Q
Y_b^2 +b_U^Q Y_t^2} \right)\,.
\ee
In this case, if $Y_b<Y_t$ then the expression in parentheses
becomes small, and we return to the small bottom Yukawa
scenario. We assume that $Y_b$ and $Y_t$ are comparable, so
that the expressions in parenthesis in Eqs.~\eqref{VQD}
and~\eqref{VQU} are $\CO$(1). We then conclude that
\be
V^{QD} \sim V^{QU} \sim V^Q \,.
\ee

The next step is to diagonalize from the left the down and up
mass matrices, thus expressing $D_L$ and $U_L$ in terms of
$V^Q\,$. Compared to Eq.~\eqref{md_left2} for $m_D\,$, we now
need to account for the fact that $F_D$ is non-universal and
not aligned with $Y_D$. Parametrizing this misalignment by the
matrix $V^D\,$, Eq.~\eqref{md_left2} is generalized to
\be
m_D m_D^\dagger \propto \rqmat\, V^Q \, \lambda_D V^D \rdsmat
V^{D\dagger} \lambda_D V^{Q\dagger} \rqmat\,,
\ee
where $r_D$ is defined as $r_Q$ but with $D \leftrightarrow Q$.
However, since the leading terms still come from $Y_b$ (so that
we can take $Y_d=Y_s=0$), $V^D$ does not play a role in this
diagonalization. Therefore, the result of
Eq.~\eqref{vq_to_vckm} holds also in the current case for the
relation between $V^Q$ and $D_L\,$, that is
\be \label{vq_to_dl}
V^Q_{12} \sim \left(D_L\right)_{12} \,, \quad V^Q_{13} \sim r_Q
\left(D_L\right)_{13} \,, \quad V^Q_{23} \sim r_Q
\left(D_L\right)_{23} \,.
\ee
Applying the same process to $m_U\,$, we see that
Eq.~\eqref{vq_to_dl} also holds after replacing $D_L \to
U_L\,$. Since $\vckm=U_L D_L^\dagger\,$, then {\it e.g.}~for
the 2-3 entry,
\be
\vckm_{ts} \sim \left(U_L\right)_{22} \left(D_L\right)_{23} +
\left(U_L\right)_{23} \left(D_L\right)_{33} \sim
\frac{V^Q_{23}}{r_Q} \,,
\ee
where the two terms in the middle are similar in magnitude,
but have different phases in general. Thus, we took only one of
them as representing the sum (omitting an order 1 correction
and the unknown phase). The bottom line is that the relations
of Eq.~\eqref{vq_to_vckm} apply to this case as well, and we
have $D_L \sim U_L \sim \vckm$.

Before continuing, it should be noted that the mass relations
in Eq.~\eqref{md_relations} are slightly changed to
\be
m_{d,s,b} \cong \frac{v}{\sqrt2} f_{Q^{1,1,3}} Y_{d,s,b}
f_{D^{1,1,3}} \,,
\ee
to include the different $c_{D^3}\,$.

In order to estimate $D_R\,$, we first need to relate $V^D$ to
an already known matrix. In the basis where $Y_D$ is diagonal,
$V^D$ diagonalizes $C_D\,$, written as
\be
C_D=a_D\cdot \mathbf{1}_3+b_D\, \lambda_D^2+ d_{DU} \lambda_D
V^{Q\dagger} \lambda_U^2 V^Q \lambda_D + \dots \,,
\ee
considering the relevant leading terms from
Eq.~\eqref{bulk_masses_long}. The mixing angles of $V^D$ are
then given by
\be
\begin{split}
V^D_{12}& \sim \frac{Y_d}{Y_s} Y_t^2 \, V^Q_{13} V^Q_{23}
\left(\frac{d_{DU}}{b_D}\right) \sim r_Q^3 r_D \frac{m_d}{m_s}
Y_t^2 \vckm_{td} \vckm_{ts}\,, \\ V^D_{13} &\sim
\frac{Y_d}{Y_b} V^Q_{13} \left( \frac{d_{DU} Y_t^2}{b_D+d_{DU}
Y_t^2} \right) \sim r_Q^2 r_D \frac{m_d}{m_b} \vckm_{td}\,, \\
V^D_{23} &\sim \frac{Y_s}{Y_b} V^Q_{23} \left( \frac{d_{DU}
Y_t^2}{b_D+d_{DU} Y_t^2} \right) \sim r_Q^2 r_D \frac{m_s}{m_b}
\vckm_{ts}\,,
\end{split}
\ee
where we again assume that the expressions in parentheses are
$\CO$(1). Now we can generalize Eq.~\eqref{md_right}:
\be
m_D^\dagger m_D \sim \rdmat\, V^{D\dagger} \, \lambda_D\,
V^{Q\dagger} \rqsmat\, V^Q \, \lambda_D \, V^D \, \rdmat \,,
\ee
and obtain $D_R\,$,
\beq
\left(D_R\right)_{12}& \sim &(r_Q^2-1)\, \frac{Y_d}{Y_s} \,
V^Q_{13}\,V^Q_{23}+V^D_{12} \sim (r_Q^2-1)r_Q^2\,
\frac{m_d}{m_s} \, \vckm_{td}\, \vckm_{ts}+ r_Q^3 r_D
\frac{m_d}{m_s} Y_t^2 \vckm_{td} \vckm_{ts}\,, \nonumber \\
\left(D_R\right)_{13} &\sim& \frac{r_Q^2-1}{r_Q^2 r_D}\,
\frac{Y_d}{Y_b}\, V^Q_{13}+ \frac{V^D_{13}}{r_D} \sim (r_Q^2-1)
\, \frac{m_d}{m_b} \, \vckm_{td}+ r_Q^2 \, \frac{m_d}{m_b} \,
\vckm_{td}\,, \\ \left(D_R\right)_{23} &\sim&
\frac{r_Q^2-1}{r_Q^2 r_D}\, \frac{Y_s}{Y_b}\, V^Q_{23}+
\frac{V^D_{23}}{r_D} \sim  (r_Q^2-1) \, \frac{m_s}{m_b} \,
\vckm_{ts}+ r_Q^2 \, \frac{m_s}{m_b} \, \vckm_{ts}\,. \nonumber
\eeq
Comparing this to the small bottom Yukawa result in
Eq.~\eqref{dr_small}, we get here for each of the angles the
same term plus an additional one, which is of the same order.
Since there is a general phase between them, we should just
take one of them as the result, so that overall there is an
$\CO$(1) correction and an undetermined phase compared to
Eq.~\eqref{dr_small}, as expected. Therefore, we are justified
in using Eq.~\eqref{fq_fd_13} in our estimates.



\begin{thebibliography}{plain}


\bibitem{Randall:1999ee}
  L.~Randall and R.~Sundrum,
  Phys.\ Rev.\ Lett.\  {\bf 83}, 3370 (1999)
  [arXiv:hep-ph/9905221].

\bibitem{ArkaniHamed:1999dc}
  N.~Arkani-Hamed and M.~Schmaltz,
  Phys.\ Rev.\  D {\bf 61}, 033005 (2000)
  [arXiv:hep-ph/9903417].

\bibitem{Grossman:1999ra}
  Y.~Grossman and M.~Neubert,
  Phys.\ Lett.\  B {\bf 474}, 361 (2000)
  [arXiv:hep-ph/9912408].

\bibitem{Agashe:2004ay}
  K.~Agashe, G.~Perez and A.~Soni,
  Phys.\ Rev.\ Lett.\  {\bf 93}, 201804 (2004)
  [arXiv:hep-ph/0406101].

\bibitem{Agashe:2004cp}
  K.~Agashe, G.~Perez and A.~Soni,
  Phys.\ Rev.\  D {\bf 71}, 016002 (2005)
  [arXiv:hep-ph/0408134].

\bibitem{Cacciapaglia:2007fw}
  G.~Cacciapaglia, C.~Csaki, J.~Galloway, G.~Marandella, J.~Terning and A.~Weiler,
  JHEP {\bf 0804}, 006 (2008)
  [arXiv:0709.1714 [hep-ph]].

\bibitem{Bona:2007vi}
  M.~Bona {\it et al.}  [UTfit Collaboration],
  arXiv:0707.0636 [hep-ph].

\bibitem{Davidson:2007si}
  S.~Davidson, G.~Isidori and S.~Uhlig,
  Phys.\ Lett.\  B {\bf 663}, 73 (2008)
  [arXiv:0711.3376 [hep-ph]].

\bibitem{Csaki:2008zd}
  C.~Csaki, A.~Falkowski and A.~Weiler,
  JHEP {\bf 0809}, 008 (2008)
  [arXiv:0804.1954 [hep-ph]].

\bibitem{Blanke:2008zb}
  M.~Blanke, A.~J.~Buras, B.~Duling, S.~Gori and A.~Weiler,
  JHEP {\bf 0903}, 001 (2009)
  [arXiv:0809.1073 [hep-ph]].

\bibitem{Bauer:2009cf}
  M.~Bauer, S.~Casagrande, U.~Haisch and M.~Neubert,
  arXiv:0912.1625 [hep-ph].

\bibitem{Cheung:2007bu}
  C.~Cheung, A.~L.~Fitzpatrick and L.~Randall,
  JHEP {\bf 0801}, 069 (2008)
  [arXiv:0711.4421 [hep-th]].

\bibitem{Huber:2003tu}
  S.~J.~Huber,
  Nucl.\ Phys.\  B {\bf 666}, 269 (2003)
  [arXiv:hep-ph/0303183].

\bibitem{Burdman:2003nt}
  G.~Burdman,
  Phys.\ Lett.\  B {\bf 590}, 86 (2004)
  [arXiv:hep-ph/0310144].

\bibitem{Moreau:2006np}
  G.~Moreau and J.~I.~Silva-Marcos,
  JHEP {\bf 0603}, 090 (2006)
  [arXiv:hep-ph/0602155].

\bibitem{Agashe:2006wa}
  K.~Agashe, G.~Perez and A.~Soni,
  Phys.\ Rev.\  D {\bf 75}, 015002 (2007)
  [arXiv:hep-ph/0606293].

\bibitem{Casagrande:2008hr}
  S.~Casagrande, F.~Goertz, U.~Haisch, M.~Neubert and T.~Pfoh,
  JHEP {\bf 0810}, 094 (2008)
  [arXiv:0807.4937 [hep-ph]].

\bibitem{Agashe:2008uz}
  K.~Agashe, A.~Azatov and L.~Zhu,
  Phys.\ Rev.\  D {\bf 79}, 056006 (2009)
  [arXiv:0810.1016 [hep-ph]].

\bibitem{Blanke:2008yr}
  M.~Blanke, A.~J.~Buras, B.~Duling, K.~Gemmler and S.~Gori,
  JHEP {\bf 0903}, 108 (2009)
  [arXiv:0812.3803 [hep-ph]].

\bibitem{Csaki:2009bb}
  C.~Csaki and D.~Curtin,
  Phys.\ Rev.\  D {\bf 80}, 015027 (2009)
  [arXiv:0904.2137 [hep-ph]].

\bibitem{Buras:2009ka}
  A.~J.~Buras, B.~Duling and S.~Gori,
  JHEP {\bf 0909}, 076 (2009)
  [arXiv:0905.2318 [hep-ph]].

\bibitem{Gedalia:2009ws}
  O.~Gedalia, G.~Isidori and G.~Perez,
  Phys.\ Lett.\  B {\bf 682}, 200 (2009)
  [arXiv:0905.3264 [hep-ph]].

\bibitem{Agashe:2004rs}
  K.~Agashe, R.~Contino and A.~Pomarol,
  Nucl.\ Phys.\  B {\bf 719}, 165 (2005)
  [arXiv:hep-ph/0412089].

\bibitem{Agashe:2005dk}
  K.~Agashe and R.~Contino,
  Nucl.\ Phys.\  B {\bf 742}, 59 (2006)
  [arXiv:hep-ph/0510164].

\bibitem{Panico:2008bx}
  G.~Panico, E.~Pont\'on, J.~Santiago and M.~Serone,
  Phys.\ Rev.\  D {\bf 77}, 115012 (2008)
  [arXiv:0801.1645 [hep-ph]].

\bibitem{Agashe:2003zs}
  K.~Agashe, A.~Delgado, M.~J.~May and R.~Sundrum,
  JHEP {\bf 0308} (2003) 050
  [arXiv:hep-ph/0308036].

\bibitem{Carena:2006bn}
  M.~S.~Carena, E.~Pont\'on, J.~Santiago and C.~E.~M.~Wagner,
  Nucl.\ Phys.\  B {\bf 759}, 202 (2006)
  [arXiv:hep-ph/0607106].

\bibitem{Cacciapaglia:2004jz}
  G.~Cacciapaglia, C.~Csaki, C.~Grojean and J.~Terning,
  Phys.\ Rev.\  D {\bf 70}, 075014 (2004)
  [arXiv:hep-ph/0401160].

\bibitem{Foadi:2004ps}
  R.~Foadi, S.~Gopalakrishna and C.~Schmidt,
  Phys.\ Lett.\  B {\bf 606}, 157 (2005)
  [arXiv:hep-ph/0409266].

\bibitem{Chivukula:2005bn}
  R.~S.~Chivukula, E.~H.~Simmons, H.~J.~He, M.~Kurachi and M.~Tanabashi,
  Phys.\ Rev.\  D {\bf 71}, 115001 (2005)
  [arXiv:hep-ph/0502162].

\bibitem{Dawson:2007yk}
  S.~Dawson and C.~B.~Jackson,
  Phys.\ Rev.\  D {\bf 76}, 015014 (2007)
  [arXiv:hep-ph/0703299].

\bibitem{Accomando:2008jh}
  E.~Accomando, S.~De Curtis, D.~Dominici and L.~Fedeli,
  Phys.\ Rev.\  D {\bf 79}, 055020 (2009)
  [arXiv:0807.5051 [hep-ph]].

\bibitem{Dawson:2008as}
  S.~Dawson and C.~B.~Jackson,
  Phys.\ Rev.\  D {\bf 79}, 013006 (2009)
  [arXiv:0810.5068 [hep-ph]].

\bibitem{Perez:2008ee}
  G.~Perez and L.~Randall,
  JHEP {\bf 0901}, 077 (2009)
  [arXiv:0805.4652 [hep-ph]].

\bibitem{Csaki:2008qq}
  C.~Csaki, C.~Delaunay, C.~Grojean and Y.~Grossman,
  JHEP {\bf 0810}, 055 (2008)
  [arXiv:0806.0356 [hep-ph]].

\bibitem{Rattazzi:2000hs}
  R.~Rattazzi and A.~Zaffaroni,
  JHEP {\bf 0104}, 021 (2001)
  [arXiv:hep-th/0012248].

\bibitem{Davoudiasl:2000wi}
  H.~Davoudiasl, J.~L.~Hewett and T.~G.~Rizzo,
  Phys.\ Rev.\  D {\bf 63}, 075004 (2001)
  [arXiv:hep-ph/0006041].

\bibitem{Davoudiasl:2007wf}
  H.~Davoudiasl, T.~G.~Rizzo, A.~Soni,
  Phys.\ Rev.\  {\bf D77}, 036001 (2008).
  [arXiv:0710.2078 [hep-ph]].

\bibitem{Kagan:2009bn}
  A.~L.~Kagan, G.~Perez, T.~Volansky and J.~Zupan,
  Phys.\ Rev.\  D {\bf 80}, 076002 (2009)
  [arXiv:0903.1794 [hep-ph]].

\bibitem{Ligeti:2010ia}
  Z.~Ligeti, M.~Papucci, G.~Perez and J.~Zupan,
  arXiv:1006.0432 [hep-ph].

\bibitem{Dobrescu:2010rh}
  B.~A.~Dobrescu, P.~J.~Fox and A.~Martin,
  arXiv:1005.4238 [hep-ph].

\bibitem{Buras:2010mh}
  A.~J.~Buras, M.~V.~Carlucci, S.~Gori and G.~Isidori,
  arXiv:1005.5310 [hep-ph].

\bibitem{Jung:2010ik}
  M.~Jung, A.~Pich and P.~Tuzon,
  arXiv:1006.0470 [hep-ph].

\bibitem{FGB} C.~Csaki, S.~J.~Lee, G.~Perez and A.~Weiler, in
    progress.

\bibitem{Carena:2007ua}
  M.~S.~Carena, E.~Pont\'on, J.~Santiago and C.~E.~M.~Wagner,
  Phys.\ Rev.\  D {\bf 76}, 035006 (2007)
  [arXiv:hep-ph/0701055].

\bibitem{PDG2010}
  J. Erler and P. Langacker, Electroweak Model and Constraints on New Physics,
  \url{http://pdg.lbl.gov/2010/reviews/rpp2010-rev-standard-model.pdf}

\bibitem{Cacciapaglia:2006mz}
  G.~Cacciapaglia, C.~Csaki, G.~Marandella and J.~Terning,
  JHEP {\bf 0702}, 036 (2007)
  [arXiv:hep-ph/0611358].

\bibitem{Agashe:2006at}
  K.~Agashe, R.~Contino, L.~Da Rold and A.~Pomarol,
  Phys.\ Lett.\  B {\bf 641}, 62 (2006)
  [arXiv:hep-ph/0605341].

\bibitem{Contino:2003ve}
  R.~Contino, Y.~Nomura and A.~Pomarol,
  Nucl.\ Phys.\  B {\bf 671}, 148 (2003)
  [arXiv:hep-ph/0306259].

\bibitem{Fitzpatrick:2007sa}
  A.~L.~Fitzpatrick, G.~Perez and L.~Randall,
  arXiv:0710.1869 [hep-ph].

\bibitem{Csaki:2009wc}
  C.~Csaki, G.~Perez, Z.~Surujon and A.~Weiler,
  Phys.\ Rev.\  D {\bf 81}, 075025 (2010)
  [arXiv:0907.0474 [hep-ph]].

\bibitem{Chivukula:1987py}
  R.~S.~Chivukula and H.~Georgi,
  Phys.\ Lett.\  B {\bf 188}, 99 (1987).

\bibitem{Hall:1990ac}
  L.~J.~Hall and L.~Randall,
  Phys.\ Rev.\ Lett.\  {\bf 65}, 2939 (1990).

\bibitem{Gabrielli:1994ff}
  E.~Gabrielli and G.~F.~Giudice,
  Nucl.\ Phys.\  B {\bf 433}, 3 (1995)
  [Erratum-ibid.\  B {\bf 507}, 549 (1997)]
  [arXiv:hep-lat/9407029].

\bibitem{Ali:1999we}
  A.~Ali and D.~London,
  Eur.\ Phys.\ J.\  C {\bf 9}, 687 (1999)
  [arXiv:hep-ph/9903535].

\bibitem{Buras:2000dm}
  A.~J.~Buras, P.~Gambino, M.~Gorbahn, S.~Jager and L.~Silvestrini,
  Phys.\ Lett.\  B {\bf 500}, 161 (2001)
  [arXiv:hep-ph/0007085].

\bibitem{D'Ambrosio:2002ex}
  G.~D'Ambrosio, G.~F.~Giudice, G.~Isidori and A.~Strumia,
  Nucl.\ Phys.\  B {\bf 645}, 155 (2002)
  [arXiv:hep-ph/0207036].

\bibitem{Buras:2003jf}
  A.~J.~Buras,
  Acta Phys.\ Polon.\  B {\bf 34}, 5615 (2003)
  [arXiv:hep-ph/0310208].

\bibitem{Buras:2005xt}
  A.~J.~Buras,
  arXiv:hep-ph/0505175.

\bibitem{Hurth:2008jc}
  T.~Hurth, G.~Isidori, J.~F.~Kamenik and F.~Mescia,
  Nucl.\ Phys.\  B {\bf 808}, 326 (2009)
  [arXiv:0807.5039 [hep-ph]].

\bibitem{Isidori:2009px}
  G.~Isidori,
  PoS E {\bf FT09}, 034 (2009)
  [arXiv:0908.0404 [hep-ph]].

\bibitem{:2003ih}
    [The LEP Collaborations: ALEPH Collaboration, DELPHI Collaboration, L3 Collaboration, OPAL Collaboration,
    the LEP Electroweak Working Group, the SLD Electroweak, Heavy Flavour Groups],
  arXiv:hep-ex/0312023.

\bibitem{:2005ema}
    [The ALEPH Collaboration, the DELPHI Collaboration, the L3 Collaboration,
    the OPAL Collaboration, the SLD Collaboration, the LEP Electroweak Working Group,
    the SLD electroweak, heavy flavour groups],
  Phys.\ Rept.\  {\bf 427}, 257 (2006)
  [arXiv:hep-ex/0509008].

\bibitem{Bouchart:2008vp}
  C.~Bouchart and G.~Moreau,
  Nucl.\ Phys.\  B {\bf 810}, 66 (2009)
  [arXiv:0807.4461 [hep-ph]].

\bibitem{Davoudiasl:2009cd}
  H.~Davoudiasl, S.~Gopalakrishna, E.~Pont\'on and J.~Santiago,
  New J.\ Phys.\  {\bf 12}, 075011 (2010)
  [arXiv:0908.1968 [hep-ph]].

\bibitem{Bouchart:2009vq}
  C.~Bouchart and G.~Moreau,
  Phys.\ Rev.\  D {\bf 80}, 095022 (2009)
  [arXiv:0909.4812 [hep-ph]].

\bibitem{Casagrande:2010si}
  S.~Casagrande, F.~Goertz, U.~Haisch, M.~Neubert and T.~Pfoh,
  arXiv:1005.4315 [hep-ph].

\bibitem{Han:2004az}
  Z.~Han and W.~Skiba,
  Phys.\ Rev.\  D {\bf 71}, 075009 (2005)
  [arXiv:hep-ph/0412166].

\bibitem{Han:2005pr}
  Z.~Han,
  Phys.\ Rev.\  D {\bf 73}, 015005 (2006)
  [arXiv:hep-ph/0510125].

\bibitem{Barbieri:2007bh}
  R.~Barbieri, B.~Bellazzini, V.~S.~Rychkov and A.~Varagnolo,
  Phys.\ Rev.\  D {\bf 76}, 115008 (2007)
  [arXiv:0706.0432 [hep-ph]].

\bibitem{Contino:2010rs}
  R.~Contino,
  arXiv:1005.4269 [hep-ph].

\bibitem{Anastasiou:2009rv}
  C.~Anastasiou, E.~Furlan and J.~Santiago,
  Phys.\ Rev.\  D {\bf 79}, 075003 (2009)
  [arXiv:0901.2117 [hep-ph]].

\bibitem{NumericalRecipes} W.~Press, S.~Teukolsky, W.
    Vetterling and B. Flannery, Numerical Recipes in C (Cambridge University Press, Cambridge, England,
    1992), 2nd ed.

\bibitem{Randall:2001gb}
  L.~Randall and M.~D.~Schwartz,
  JHEP {\bf 0111}, 003 (2001)
  [arXiv:hep-th/0108114].

\bibitem{Pomarol:2000hp}
  A.~Pomarol,
  Phys.\ Rev.\ Lett.\  {\bf 85}, 4004 (2000)
  [arXiv:hep-ph/0005293].

\bibitem{Randall:2001gc}
  L.~Randall and M.~D.~Schwartz,
  Phys.\ Rev.\ Lett.\  {\bf 88}, 081801 (2002)
  [arXiv:hep-th/0108115].

\bibitem{Agashe:2004ci}
  K.~Agashe and G.~Servant,
  Phys.\ Rev.\ Lett.\  {\bf 93}, 231805 (2004)
  [arXiv:hep-ph/0403143].

\bibitem{Agashe:2004bm}
  K.~Agashe and G.~Servant,
  JCAP {\bf 0502}, 002 (2005)
  [arXiv:hep-ph/0411254].

\bibitem{Agashe:2005vg}
  K.~Agashe, R.~Contino and R.~Sundrum,
  Phys.\ Rev.\ Lett.\  {\bf 95}, 171804 (2005)
  [arXiv:hep-ph/0502222].

\bibitem{Agashe:2009ja}
  K.~Agashe, K.~Blum, S.~J.~Lee and G.~Perez,
  Phys.\ Rev.\  D {\bf 81}, 075012 (2010)
  [arXiv:0912.3070 [hep-ph]].

\bibitem{Cheng:2002iz}
  H.~C.~Cheng, K.~T.~Matchev and M.~Schmaltz,
  Phys.\ Rev.\  D {\bf 66}, 036005 (2002)
  [arXiv:hep-ph/0204342].

\bibitem{Barbieri:1987fn}
  R.~Barbieri and G.~F.~Giudice,
  Nucl.\ Phys.\  B {\bf 306}, 63 (1988).

\bibitem{Anderson:1994dz}
  G.~W.~Anderson and D.~J.~Castano,
  Phys.\ Lett.\  B {\bf 347}, 300 (1995)
  [arXiv:hep-ph/9409419].

\bibitem{Gedalia:2010rj}
  O.~Gedalia and G.~Perez,
  arXiv:1005.3106 [hep-ph].

\bibitem{Gedalia:2009kh}
  O.~Gedalia, Y.~Grossman, Y.~Nir and G.~Perez,
  Phys.\ Rev.\  D {\bf 80}, 055024 (2009)
  [arXiv:0906.1879 [hep-ph]].

\bibitem{Buras:2010pz}
  A.~J.~Buras, D.~Guadagnoli and G.~Isidori,
  Phys.\ Lett.\  B {\bf 688}, 309 (2010)
  [arXiv:1002.3612 [hep-ph]].

\bibitem{Ellis:2007kb}
  J.~R.~Ellis, J.~S.~Lee and A.~Pilaftsis,
  Phys.\ Rev.\  D {\bf 76}, 115011 (2007)
  [arXiv:0708.2079 [hep-ph]].

\bibitem{Colangelo:2008qp}
  G.~Colangelo, E.~Nikolidakis and C.~Smith,
  Eur.\ Phys.\ J.\  C {\bf 59}, 75 (2009)
  [arXiv:0807.0801 [hep-ph]].

\bibitem{Mercolli:2009ns}
  L.~Mercolli and C.~Smith,
  Nucl.\ Phys.\  B {\bf 817}, 1 (2009)
  [arXiv:0902.1949 [hep-ph]].

\bibitem{Paradisi:2009ey}
  P.~Paradisi and D.~M.~Straub,
  Phys.\ Lett.\  B {\bf 684}, 147 (2010)
  [arXiv:0906.4551 [hep-ph]].

\bibitem{Gedalia:2010zs}
  O.~Gedalia, L.~Mannelli and G.~Perez,
  arXiv:1002.0778 [hep-ph].

\bibitem{Gedalia:2010mf}
  O.~Gedalia, L.~Mannelli and G.~Perez,
  arXiv:1003.3869 [hep-ph].

\bibitem{Abazov:2010hv}
  V.~M.~Abazov {\it et al.}  [D0 Collaboration],
  arXiv:1005.2757 [hep-ex].

\bibitem{Abazov:2008fj}
  V.~M.~Abazov {\it et al.}  [D0 Collaboration],
  Phys.\ Rev.\ Lett.\  {\bf 101}, 241801 (2008)
  [arXiv:0802.2255 [hep-ex]].

\bibitem{CDF:2010}
  L.~Oakes [CDF Collaboration],
  talk at FPCP 2010, May 25-29, Torino, Italy,
  \url{http://agenda.infn.it/getFile.py/access?contribId=12&resId=0&materialId=slides&confId=2635}.

\bibitem{Eberhardt:2010bm}
  O.~Eberhardt, A.~Lenz and J.~Rohrwild,
  arXiv:1005.3505 [hep-ph].

\bibitem{Dighe:2010nj}
  A.~Dighe, A.~Kundu and S.~Nandi,
  arXiv:1005.4051 [hep-ph].

\bibitem{Chen:2010wv}
  C.~H.~Chen and G.~Faisel,
  arXiv:1005.4582 [hep-ph].

\bibitem{Bauer:2010dg}
  C.~W.~Bauer and N.~D.~Dunn,
  arXiv:1006.1629 [hep-ph].

\bibitem{Deshpande:2010hy}
  N.~G.~Deshpande, X.~G.~He and G.~Valencia,
  arXiv:1006.1682 [hep-ph].

\bibitem{Batell:2010qw}
  B.~Batell and M.~Pospelov,
  arXiv:1006.2127 [hep-ph].

\bibitem{Kurachi:2010fa}
  M.~Kurachi and T.~Onogi,
  arXiv:1006.3414 [hep-ph].

\bibitem{Chen:2010aq}
  C.~H.~Chen, C.~Q.~Geng and W.~Wang,
  arXiv:1006.5216 [hep-ph].

\bibitem{Parry:2010ce}
  J.~K.~Parry,
  arXiv:1006.5331 [hep-ph].

\bibitem{Randall:1998te}
  L.~Randall and S.~f.~Su,
  Nucl.\ Phys.\  B {\bf 540}, 37 (1999)
  [arXiv:hep-ph/9807377].

\bibitem{Isidori:2010kg}
  G.~Isidori, Y.~Nir and G.~Perez,
  arXiv:1002.0900 [Unknown].

\bibitem{Agashe:2009di}
  K.~Agashe and R.~Contino,
  Phys.\ Rev.\  D {\bf 80}, 075016 (2009)
  [arXiv:0906.1542 [hep-ph]].

\bibitem{Azatov:2009na}
  A.~Azatov, M.~Toharia and L.~Zhu,
  Phys.\ Rev.\  D {\bf 80}, 035016 (2009)
  [arXiv:0906.1990 [hep-ph]].

\bibitem{Duling:2009pj}
  B.~Duling,
  JHEP {\bf 1005}, 109 (2010)
  [arXiv:0912.4208 [hep-ph]].

\bibitem{Agashe:2006hk}
  K.~Agashe, A.~Belyaev, T.~Krupovnickas, G.~Perez and J.~Virzi,
  Phys.\ Rev.\  D {\bf 77}, 015003 (2008)
  [arXiv:hep-ph/0612015].

\bibitem{Lillie:2007yh}
  B.~Lillie, L.~Randall and L.~T.~Wang,
  JHEP {\bf 0709}, 074 (2007)
  [arXiv:hep-ph/0701166].

\bibitem{TevatronElectroweakWorkingGroup:2010yx}
  Tevatron Electroweak Working Group, CDF~Collaboration and D0~Collaboration,
  arXiv:1007.3178 [hep-ex].

\bibitem{:2009nu}
  Tevatron Electroweak Working Group, CDF~Collaboration and D0~Collaboration,
  arXiv:0908.1374 [hep-ex].

\bibitem{Carena:2003fx}
  M.~S.~Carena, A.~Delgado, E.~Pont\'on, T.~M.~P.~Tait and C.~E.~M.~Wagner,
  Phys.\ Rev.\  D {\bf 68}, 035010 (2003)
  [arXiv:hep-ph/0305188].

\bibitem{Grojean:2006nn}
  C.~Grojean, W.~Skiba and J.~Terning,
  Phys.\ Rev.\  D {\bf 73}, 075008 (2006)
  [arXiv:hep-ph/0602154].

\bibitem{Cacciapaglia:2006pk}
  G.~Cacciapaglia, C.~Csaki, G.~Marandella and A.~Strumia,
  Phys.\ Rev.\  D {\bf 74}, 033011 (2006)
  [arXiv:hep-ph/0604111].

\bibitem{Goldberger:2002cz}
  W.~D.~Goldberger and I.~Z.~Rothstein,
  Phys.\ Rev.\ Lett.\  {\bf 89}, 131601 (2002)
  [arXiv:hep-th/0204160].

\bibitem{Agashe:2002bx}
  K.~Agashe, A.~Delgado and R.~Sundrum,
  Nucl.\ Phys.\  B {\bf 643}, 172 (2002)
  [arXiv:hep-ph/0206099].

\bibitem{Contino:2002kc}
  R.~Contino, P.~Creminelli and E.~Trincherini,
  JHEP {\bf 0210}, 029 (2002)
  [arXiv:hep-th/0208002].

\bibitem{Goldberger:2002hb}
  W.~D.~Goldberger and I.~Z.~Rothstein,
  Phys.\ Rev.\  D {\bf 68}, 125011 (2003)
  [arXiv:hep-th/0208060].

\bibitem{Choi:2002ps}
  K.~w.~Choi and I.~W.~Kim,
  Phys.\ Rev.\  D {\bf 67}, 045005 (2003)
  [arXiv:hep-th/0208071].

\bibitem{Agashe:2002pr}
  K.~Agashe, A.~Delgado and R.~Sundrum,
  Annals Phys.\  {\bf 304}, 145 (2003)
  [arXiv:hep-ph/0212028].

\bibitem{Goldberger:2003mi}
  W.~D.~Goldberger and I.~Z.~Rothstein,
  Phys.\ Rev.\  D {\bf 68}, 125012 (2003)
  [arXiv:hep-ph/0303158].

\bibitem{Lavoura:1992np}
  L.~Lavoura and J.~P.~Silva,
  Phys.\ Rev.\  D {\bf 47}, 2046 (1993).

\bibitem{Xing:2007fb}
  Z.~z.~Xing, H.~Zhang and S.~Zhou,
  Phys.\ Rev.\  D {\bf 77}, 113016 (2008)
  [arXiv:0712.1419 [hep-ph]].

\bibitem{Bamert:1996px}
  P.~Bamert, C.~P.~Burgess, J.~M.~Cline, D.~London and E.~Nardi,
  Phys.\ Rev.\  D {\bf 54}, 4275 (1996)
  [arXiv:hep-ph/9602438].

\end{thebibliography}
\end{document}